\documentstyle[11pt,aaspp4]{article}

\lefthead{Woods and Fahlman}
\righthead{Correlation Function for Faint Galaxies}

\begin{document}

\title{Measuring the Angular Correlation Function for Faint Galaxies in \\
High Galactic Latitude Fields}

\author{David Woods\altaffilmark{1} and Gregory G. Fahlman\altaffilmark{1}}
\affil{Dept. of Physics and Astronomy, 2219 Main Mall, 
University of British Columbia, Vancouver, B.C. V6T 1Z4}

\altaffiltext{1}{Visiting Astronomer, Canada--France--Hawai'i Telescope
(CFHT), operated by the National Research Council of Canada, le Centre
National de la Recherche Scientifique de France, and the University of
Hawai'i.} 

\begin{abstract}

A photometric survey of faint galaxies in three high Galactic 
latitude fields (each $\sim49~\rm{arcmin^{2}}$) with sub-arcsecond seeing 
is used to study the clustering properties of 
the faint galaxy population.  
Multi-color 
photometry of the galaxies has been obtained to magnitude 
limits of $V\sim25$, $R\sim25$ and $I\sim24$.  
Angular correlation analysis is applied to magnitude-limited and 
color-selected samples of 
galaxies from 
the three fields for angular separations ranging from $10-126''$.  
General agreement is obtained with 
other recent studies which show that the amplitude of the angular 
correlation function, 
$\omega(\theta)$, is smoothly 
decreasing as a 
function of limiting magnitude.  
The observed decline of $\omega(\theta)$ 
rules out the viability of ``maximal merger'' galaxy evolution 
models.  

Using 
redshift distributions extrapolated to faint magnitude limits, 
models of galaxy clustering evolution are 
calculated and compared to the observed $I$-band $\omega(\theta)$.  
Faint galaxies are determined to have correlation lengths and 
clustering evolution parameters of either $r_{0}\sim4~h^{-1}~{\rm Mpc}$ and 
$\epsilon\sim0-1$; $r_{0}\sim5-6~h^{-1}~{\rm Mpc}$ and $\epsilon>1$; or 
$r_{0}\sim2-3~h^{-1}~{\rm Mpc}$ and $\epsilon\sim-1.2$, 
assuming $q_{0}=0.5$ and with 
$h=H_{0}/100~{\rm km~s}^{-1}~{\rm Mpc}^{-1}$.  
The latter 
case is for clustering fixed in co-moving coordinates and is probably
unrealistic since most local galaxies are observed to be 
more strongly clustered.  Even though  
the first of the three cases has the most reasonable rate of clustering 
evolution, distinguishing the correct $r_{0}$ for the faint 
galaxies is not possible with the current data.  
No significant variations in the
clustering amplitude as a function of color 
are detected, for all the color-selected galaxy samples considered.  
The validity of this result is discussed in relation to 
other determinations of $\omega(\theta)$ for galaxies selected by color.  
 
\end{abstract}


\section{Introduction}

Angular correlation function analysis of large photometric samples of 
galaxies is a popular tool for 
quantifying the large-scale structure of the local universe.  
Its power as a diagnostic of the galaxy distribution lies in the 
simplicity of its application, essentially requiring only a counting 
of pairs of galaxies at given angular separations and normalizing 
these results with respect to the number of pairs expected for a random
distribution.  The angular correlation function, $\omega(\theta)$, 
is a two-dimensional analogue of the spatial two-point correlation function, 
$\xi(r)$, where the latter quantity 
has the important property of being the Fourier
transform of the power spectrum, $P(k)$, of the galaxy distribution.  
Measurements of either $\xi(r)$ or $P(k)$ are important 
tests of structure 
formation models, 
such as Cold Dark Matter (CDM) scenarios (Davis {\it et al.}\markcite{D85} 1985), 
and are also necessary for 
understanding the relationship between nearby, bright galaxies and 
their faint, distant counterparts, along with the evolutionary 
processes that link the two populations.  

The angular and spatial 
two-point correlation functions for bright magnitude-limited samples 
of nearby galaxies have been determined, in various forms, by several authors 
over the last few decades (see Peebles \markcite{P80}1980, \markcite{P93}1993, 
and references therein).  
Numerous studies have measured $\xi(r)$ to be a 
power-law, 
$(r/r_{0})^{-\gamma}$, 
with $\gamma\sim1.7-1.8$ and the correlation length, $r_{0}(z=0)$, is 
estimated to be $\sim5~h^{-1}~{\rm Mpc}$ for local galaxy populations 
(Davis \& Peebles \markcite{DP83}1983, 
Loveday {\it et al.} \markcite{L92}1992). 
There is still considerable uncertainty concerning 
the exact normalization and slope due to possible systematic errors from
morphological mixing, redshift distortions from clusters of galaxies 
and differing approaches to the correlation analysis 
(Bernstein {\it et al.} \markcite{B94b}1994, 
Loveday {\it et al.} \markcite{L95}1995).
Another potential problem for studies of nearby galaxies may be that a 
significant population of 
low surface brightness galaxies is being systematically ignored, 
due to high surface brightness selection effects, in contrast to 
faint galaxy surveys where this is not a problem 
(McGaugh \markcite{M94}1994).   

Pioneering 
photographic studies of faint galaxies 
done by Phillipps {\it et al.} \markcite{P78}(1978), 
Koo \& Szalay \markcite{KS84}(1984), 
Stevenson {\it et al.} \markcite{S85}(1985) and 
Pritchet \& Infante \markcite{PI86}(1986) determined 
$\omega(\theta)$ down to magnitude limits of $B<23-24$.  
Only in the last decade has it been possible to measure 
$\omega(\theta)$ for
fainter galaxy samples because of the emergence of CCD 
imaging cameras, in particular the large-format devices.  
Efstathiou {\it et al.} \markcite{E91}(1991, hereafter EBKTG) 
found the faint blue galaxy population 
in their CCD images (for $24<B_{J}<26$) 
to be weakly clustered at $30''$ separations 
relative to local galaxy populations.  
They 
concluded that either 
(1) most of the faint galaxies were members of 
an hitherto unobserved population which had faded away by the current
epoch, (2) galaxy clustering was insufficiently 
described by basic models of gravitational
instability or (3) that space-time geometry departed significantly
from an Einstein-de Sitter universe.  
However, as Koo \& Kron \markcite{KK92}(1992)
point out, the EBKTG\markcite{E91} 
result implicitly assumes that galaxies with
different morphologies have similar intrinsic clustering properties.  
This is not the case locally (Loveday {\it et al.} \markcite{L95}1995, Giovanelli 
{\it et al.} \markcite{G86}1986, Davis \& Geller \markcite{DG76}1976) 
and clearly this is 
an effect which needs to be addressed by using various sample selection 
criteria for faint galaxies such as
morphologies and colors, in addition to magnitude limits.  

Neuschaefer {\it et al.} \markcite{N91}(1991, hereafter NWD) 
measured $\omega(\theta)$ down to 
$g\sim25$ 
and found a similar amplitude as EBKTG\markcite{E91} at $g\sim24.8$.  The monotonic
decrease of the amplitude of $\omega(\theta)$ for a given angular separation
as a function of survey magnitude limit  
demonstrated by NWD\markcite{N91} has been
observed in a number of other studies (Pritchet \& 
Infante \markcite{PI92}1992, 
Couch {\it et al.} \markcite{C93}1993, Roche {\it et al.} \markcite{R93}1993, 
Roukema \& Peterson \markcite{RP94}1994, 
Brainerd {\it et al.} \markcite{BSM95}1995, hereafter BSM, 
and Metcalfe {\it et al.} \markcite{M95}1995, hereafter MSFR, to name a few).  
This decrease
of clustering with survey depth shows some indication of beginning 
to flatten out at the faintest limits in some studies 
(Roche {\it et al.} \markcite{R93}1993 and MSFR\markcite{M95})
but the errors involved in these measurements preclude any firm conclusions
from being made.  

Studying the clustering of galaxies with color-selected samples 
is a technique which has only recently been applied to discriminate 
faint galaxies of differing morphologies 
(Bernstein {it et al.} \markcite{B94b}(1994), hereafter BTBJ; 
Landy, Szalay \& Koo \markcite{LSK96}1996, 
hereafter LSK; Roche {\it et al.} \markcite{R96}1996).  
Using $U-R_{F}$ as the color discriminant, 
LSK\markcite{LSK96} 
found a factor of $\gtrsim10$ increase in the amplitude
of the angular correlation function for the extremely 
blue {\it and} red subsets of galaxies 
with $20<B_{J}<23.5$ 
in their sample of 5900 galaxies.  
LSK\markcite{LSK96} 
claim the $\omega(\theta)$ excess for the reddest galaxies is due to 
intrinsic clustering of the sample (morphology-density relation) since
most of these objects are probably E/S0 galaxies and both fields contain
known clusters.  The increase of the clustering amplitude for the 
bluest objects in the sample is explained 
as being caused by a faint population of 
galaxies with $z<0.3$.  Roche {\it et al.} \markcite{R96}(1996) calculated 
$\omega(\theta)$ for color-selected samples to significantly 
fainter magnitude limits ($B\simeq25.5$ and $R\simeq24.5$) than 
LSK\markcite{LSK96}.  
For $\sim7000$ galaxies, they 
determined the amplitude of $\omega(\theta)$ for the red ($(B-R)>1.5$) 
sample to be higher than that calculated for the blue ($(B-R)<1.5$) galaxies.  
This result led Roche {\it et al.}\markcite{R96} to suggest that the decrease in 
the amplitude of $\omega(\theta)$ for all galaxies 
with $B>23$ is caused by the same blue
galaxies that are responsible for the number counts excess around this
magnitude range.  Using a pure luminosity evolution model they 
explain the correlation function color dependence at $B\sim25$ 
as being due to red galaxies with $z<1$, in addition to 
the blue galaxy sample consisting
of both late-type dwarfs at low/moderate redshifts and evolving
$L^{*}$ galaxies having redshifts from $z\sim0.5-3$.  

In this paper, $\omega(\theta)$ is calculated for a sample 
of galaxies imaged in $V$, $R$ and $I$ 
to respective magnitude limits of 
25, 25 and 24, combining data from 
three different high Galactic latitude fields.
Woods {\it et al.} \markcite{W95}(1995) determined 
the close pair fraction for one of the 
fields (NF1) studied here.  
This nearest neighbour approach 
is complementary to the correlation analysis since it 
measures clustering behaviour at the smallest possible angular separations 
while the greater numbers of galaxies analysed in 
this paper can be used to  
estimate the angular correlation function
over a range of larger angular separations 
(as an aside, note that equations 1 and 2 in 
Woods {\it et al.} \markcite{W95}1995 are both missing a factor of 
$\rho$ in the integrand).  
Galaxy samples presented here are among
the deepest yet used for 
measurements of $\omega(\theta)$.  
Also, the multi-bandpass 
data allows the clustering variations with color to be 
measured for these faint galaxies.  
Observations and preliminary data reduction and analysis techniques 
are outlined in \S\S 2 and 3, respectively.  
The adopted approach for estimating the angular correlation function, 
along with a summary of the galaxy clustering 
model used, is presented in \S 4.  Clustering results for 
magnitude-limited and
color-selected samples of galaxies are given in \S 5.  
This section also contains a comparison of the clustering detected in the 
$I$-band to models of the {\it spatial} correlation 
function, which are calculated with extrapolated redshift distributions 
provided by the CFRS (Lilly {\it et al.} \markcite{L95a}1995a).  
Finally, possible interpretations of the results are discussed and 
summarized 
in \S\S 6 and 7.

\section{Observations}

The $V$, $R$ and $I$ images used in this study were obtained at the prime focus
of the Canada-France-Hawai'i telescope using FOCAM and 
the LICK1 and LICK2 large-format, $2k~{\rm x}~2k$
CCDs between 1991 April and 1993 March.  
The image scale for both LICK devices is $0\farcs207$ per pixel 
so the full field-of-view of the CCD is $\sim7'$ on a side.  
These images of high Galactic latitude ``blank'' fields
were originally obtained for a 
survey of Population II halo stars 
(see Richer \& Fahlman \markcite{RF92}1992) 
but are also useful for studying properties of the faint galaxy 
population.  Three fields with {\it north} Galactic latitudes were 
observed and 
are dubbed NF1, NF2 and NF3.  
The right ascension and declination, and the corresponding Galactic 
coordinates, of the centers of these three blank fields are given in Table 1, 
along with the time of the observing runs.  
Fields were specifically chosen to have no observable objects on the 
Palomar Sky Survey photographs and a lack of any Zwicky clusters.  
Seeing for the frames used in the final summed images is uniformly excellent, 
ranging from $0\farcs5$ to 
$1''$.  Good seeing is essential for acquiring deep images in a reasonable
amount of exposure time.  The filter bandpasses used, and 
the total exposure times and average seeing for 
the summed frames in each color and field, are summarized in Table 2. 
$V+R+I$ frames which demonstrate the total multi-band exposure for 
each field are shown in Figs. 1, 2 and 3.  

\section{Data Reduction and Analysis}

\subsection{Image Pre-processing and Summation}

Exposures of the fields were obtained 
typically for 900-1200s, in between which the telescope would be 
dithered by $\simeq10''$ in the cardinal directions such that   
program frames could be used to generate a sky-flat
in each bandpass, for flat-fielding purposes.  
Various programs from the {\it IRAF~\footnote{\rm Image Reduction 
and Analysis Facility, a 
software system distributed by the National Optical Astronomy 
Observatories (NOAO).}} 
package were used to
do the pre-processing of the CCD images.  
All frames had the instrumental dc level, as 
determined from the CCD bias region, removed with the 
{\it linebias} routine and a median bias frame was then subtracted from each 
program frame to remove the pixel-to-pixel pattern.  
The pixel response function for each bandpass was determined 
by constructing a sky-flat from the median of the corresponding 
dithered program frames.  This ``superflat'' was generated using anywhere 
from $7-10$ program frames, with this set including the data 
frames used for the final combined frame.  Any artifacts caused by the 
use of a data frame in the making of its own flatfield were dealt with
by masking brighter objects, where these effects were most significant.  

After the flat-fielding step, the individual frames were registered with 
the {\it IRAF} routine {\it imalign}.   
The alignment interpolation was performed to the nearest pixel.  
There is a slight rotation ($\sim1\arcdeg$)
between the 
images obtained of NF2 in 1992 and 1993 and the {\it IRAF} routines 
{\it geomap} and {\it geotran}
were used to align these data.  
Finally, an exposure weighted average of the aligned program frames 
in each field and for each bandpass was obtained with the {\it IRAF} 
routine {\it imcombine}.  
CCD frames with poor seeing were 
not included in the average 
to avoid resolution degradation.
The final combined frames were always found to be flat to 
within $\lesssim1\%$.  Image flatness is an important feature 
of our processed images in that it allows 
accurate faint galaxy photometry to be determined.  
Cosmic rays were also removed, as a final step, using the {\it IRAF} routine 
{\it cosmicrays}.  No obvious differences in the final images 
were found when the 
cosmic rays were removed before or after the combination of
the individual data frames.  Careful selection of the detection 
threshold and other parameters for the {\it cosmicrays} routine resulted 
in the removal of most of the cosmic rays from the final combined frames.  

\subsection{Calibrations}

Standard star fields were observed on every night of each observing run.  
The fields used are summarized in Table 3, with the pertinent references.
The zero points in the calibration equations were very stable, not
varying 
more than $\sim0.2$ mag 
between observing runs.  
Color terms found in the calibration solutions
for the three runs were accounted for by small offsets ($\leq0.05$)
based on a mean color for the faint galaxies or were disregarded 
due to the calculated coefficient being negligible.  
Stetson's (\markcite{St87}1987, \markcite{St90}1990) DAOGROW and DAOPHOT programs 
were used to determine accurate aperture corrections for the standard star 
frames.  
The only departure from standard calibration techniques necessary 
was in the case of the NF2 $V$ and $I$ final frames, which were comprised of 
images from two different observing runs.  A set of $\sim20$ stars in the 
NF2 field were chosen to be secondary standards.  Magnitude offsets 
between these secondary standards photometered on a single frame 
from the 1992 June run, and the final, averaged
frames were determined.  
These small offsets ($\leq0.05$ mag) 
were found to be color independent 
and therefore could be applied directly to the magnitudes obtained 
from the final averaged $V$ and $I$ frames for NF2, to correct for the
variations between the two observing runs.  For more details of the 
calibrations the reader is referred to 
Woods \markcite{W96}(1996).  

\subsection{Generating FOCAS Catalogs}

The objects in our final CCD frames were detected and analysed using FOCAS 
(Jarvis \& Tyson \markcite{JT81}1981) routines 
with slight modifications to the
standard analysis procedures, some of which are outlined in 
Valdes (\markcite{V83}1983, \markcite{V93}1993).  
The two key input parameters for FOCAS are the detection threshold, 
given as a multiple of the sky variance ($\sigma_{sky}$), and the minimum 
area for the objects detected, in numbers of pixels.  After experimenting
with these parameters and ensuring that spurious detections were 
minimized, we adopted a threshold of $2.5\sigma_{sky}$ and a minimum
area which corresponded to the seeing disk for the poorest resolved 
frame from the three bandpasses taken of a given field 
(also see Steidel \& Hamilton \markcite{SH93}1993).  
Since we use conservative 
magnitude limits in this study (see \S 3.4) where galaxy 
incompleteness is negligible and the success of the galaxy 
detection is checked thoroughly by eye, we are confident that the 
threshold and minimum area parameters chosen are appropriate.  

Two approaches were used for the initial detection of the objects
in our fields.  The first technique was to use a
``master'' frame ($V+R+I$), where the average
frames from each bandpass were normalized to a common sky level 
in counts (Smail {\it et al.} \markcite{Sm94}1994).  An object list was generated 
from the master frame detections, then the galaxy 
magnitudes were measured off of the average images in each respective filter.  
This works quite well for 
fields where data have been obtained with comparable magnitude limits in  
the three filters (NF1).  Note that $V \sim R+0.5$ and $R \sim I+0.75$ 
(see galaxy color histograms in Fig. 7) 
for galaxies with late-type morphologies 
within the redshift regime that approximately corresponds
to our magnitude limits (Frei \& Gunn \markcite{FG94}1994).  
Exceptions to this are ellipticals 
which become harder to detect since the $4000$\AA~ 
break 
has been redshifted beyond the V filter at $z\geq0.5$.
In fields where the faint limits
were not roughly equivalent 
from filter-to-filter (NF2 and NF3), the second approach was to 
have the initial detection
of the objects done in each individual bandpass.  The objects 
found in each filter were then
matched in master catalogs in order to provide color information.  
In particular, the $R$-band data taken for NF2 and NF3 were found to be 
deeper than the $V$ and $I$ data. 

FOCAS programs are applied to either the master frame of a field or 
the averaged frame in each filter (i.e., the ``detection'' frame)
to determine an initial 
catalog of objects, with a slight modification.  
The detection algorithm
for FOCAS will find different numbers of objects, varying by a few percent, 
depending on the orientation of the frame.  
This variation in numbers is 
due to the line-by-line nature of the detection algorithm and 
the fact that the threshold for a particular line depends on the sky history
from the previous lines.  We work around this problem by rotating the detection frame
through 90 degree increments and matching the resultant four catalogs to 
produce a final catalog.  An object is included in the final catalog 
if it is detected in all the catalogs from the four 
orientations.  
We used the 
``builtin'' FOCAS filter (1 2 3 2 1) 
for convolution with the image under consideration 
during the detection process 
to reduce the number 
of spurious objects in the final catalogs.  
The final catalog is also filtered to remove detections
of objects which lie within ``masked'' areas of the frame.  Masked
regions include saturated stars, very bright galaxies, 
bad columns, 
vignetted corners and other artifacts which generate spurious detections.  
The masked area of each field is 
typically only a few percent of the 
total number of pixels in the final frame.  

Following the application of the detection algorithm, the sky
values are determined for each frame 
using the standard {\it sky} and {\it skycorrect}
routines in FOCAS.  No significant dependence of the measured sky values
on the orientation of the frame 
was found.  
Detections listed in the final catalogs were split into individual 
objects and the magnitudes were evaluated using the default FOCAS programs.  
Splitting of multi-component groups into individual galaxies (and stars)
was easily accomplished for the separations over which 
the angular correlation function is calculated, mostly due to 
the uniformly excellent seeing of the data set.  
All objects were split 
with confidence down to separations of $\sim1''$, 
as in Woods {\it et al.} \markcite{W95}(1995).  
Measurement of galaxy
magnitudes is discussed further in the following section.  A point spread
function (PSF) was determined from $\sim15-20$ bright stars found
in each field for each bandpass.  The PSF is used in the FOCAS object
classification algorithm {\it resolution}, to separate galaxies and 
stars from spurious objects in the final master catalog.  
Separation of stars from galaxies in the final samples was {\it not} done
with the FOCAS classification routine but with an approach outlined 
in \S 4.3.  Objects listed in the final catalogs were all checked by
eye to confirm their detection.  A few spurious objects remained 
at this juncture, and were removed from further analysis, but their
numbers were small relative to the final galaxy sample.  

\subsection{Final Photometry}

Total exposures for the three bandpasses, in a 
given field, 
were obtained so as to be comparable in depth 
to maximize the color information for
the greatest number of objects.  In practice this is difficult to do 
at the telescope due to varying seeing, along with 
weather and observing time constraints. 
However, we obtained fairly uniform sampling of the objects 
allowing us to determine $V$, $R$ and $I$ magnitudes for $\sim1000$ galaxies
in each field, with the exceptions of the $V$ and $I$ data for NF2.  Slightly
lower numbers of faint objects for the NF2 data were detected due to the 
data sets being collected during two observing runs which created 
a small loss in area from slightly mismatched fields.  Only the field area
which is coincident on all the data frames is included in the 
final detection frames, so that uniform magnitude-limited
samples are generated.  

Magnitudes were evaluated for each bandpass and field yielding final 
lists of isophotal and aperture magnitudes, and colors for all the objects.  
Aperture magnitudes were found to be a more reliable measure of 
the total brightness of the faint galaxies 
than isophotal magnitudes.  
Also, the isophotal magnitude limit was found to be 
unrealistically faint.  
However, the use of a small fixed aperture is not appropriate 
for the galaxies in the sample which have 
a significant angular extent.  
Hence ``hybrid'' magnitudes are adopted: an aperture of $3''$ in diameter was  
used for the galaxies with an average diameter $\leq3''$ and 
isophotal magnitudes were used for galaxies with characteristic
sizes larger than this.  
Isophotal and aperture magnitudes are in good agreement at the 
magnitude range where the transition between the two measures occurs, 
typically having differences $\lesssim0.1$ mag.  

Isophotal 
magnitudes are usually used for galaxies which are up to three magnitudes 
below the bright magnitude limit ($V\sim20-23$, $R\sim20-23$ and $I\sim19-22$) 
where galaxies are $>3''$ in mean
diameter.  
Since the majority of the galaxies in any of our samples is within two 
magnitudes of the faint magnitude limit ($V>23$, $R>23$ and $I>22$) 
the galaxies measured with isophotal magnitudes are a small contribution
to the final sample size.  
The 
isophotal magnitudes used are {\it not} the ``total'' magnitudes which 
FOCAS generates (Valdes \markcite{V83}1983) 
since we prefer to avoid 
doubling the isophotal area, from the 
initially determined isophote, for the 
flux measurements.  
Considering the small numbers of 
galaxy magnitudes measured with isophotes this choice does not 
have a significant effect on the final magnitude-limited galaxy samples.  
Apertures of $3''$ were chosen since they were determined to 
be large enough to contain most of the flux from 
the majority of the faint galaxies.  
The adopted aperture size of $3''$ corresponds to a physical 
scale of $\sim11~h^{-1}~{\rm kpc}$ and $\sim13~h^{-1}~{\rm kpc}$ for redshifts 
of $z=0.5$ and $1.0$, respectively, using 
$h=H_{0}/100~{\rm km~s}^{-1}~{\rm Mpc}^{-1}$ and $q_{0}=0.5$.  
See \S 2.3 of Lilly {\it et al.} \markcite{LCG91}(1991) for some further discussion
of the benefits of aperture photometry for faint galaxies, as opposed
to using isophotal photometry.   

Number counts, in $V$, $R$ and $I$, are given for the three fields
in Figs. 4, 5 and 6, respectively.  
Bright stars have been removed from these counts in the manner 
outlined in \S 4.3 and hybrid magnitudes 
are used to calculate these counts.  
The slopes of the galaxy counts are listed 
in Table 4.   There is good agreement between the slopes 
determined in different fields for a given bandpass.  
These are encouraging 
results since we want to calculate $\omega(\theta)$ by averaging
the galaxy clustering behaviour in NF1-3, thus requiring
field-to-field uniformity.  
To avoid crowding the plots in Figs. 4-6 with a plethora of 
data points from other studies we have compared our 
$R$ number counts normalization and slope 
to the compilation of Metcalfe {\it et al.} \markcite{M91}(1991) 
and the counts of Hudon \& Lilly \markcite{HL96}(1996, hereafter HL) 
both plotted in the latter study's Fig. 1.  
We find excellent agreement with the $R$ counts of 
HL\markcite{HL96} and Metcalfe {\it et al}\markcite{M91}.  
Checking our $V$ and $I$ data is more problematic due to 
a scarcity of published number counts in these bandpasses.  
Our $V$ counts are in good agreement with the observations shown in Fig. 2 of 
Smail {et al.} \markcite{Sm95}(1995) except we observe no break 
at $V\sim24.25$, but this is not surprising considering our magnitude limit
is $V\sim25$.  The $I$ counts from the current study have a 
slightly steeper slope 
and different normalization than the Smail {\it et al.}\markcite{Sm95} 
data but are consistent with the Lilly {\it et al.} \markcite{LCG91}(1991) 
number counts.  The source of the difference with Smail {\it et al.}\markcite{Sm95} 
is not clear, however,  
they suggest that the discrepancy could be 
due to other workers choosing fields which are devoid of bright galaxies 
therefore biasing the count slope to be steeper.  
Nevertheless, we find our number counts to be in generally good agreement 
with other groups' observations.  

From the number counts it is easily seen that conservative 
magnitude limits for the data are $V\sim25$, $R\sim25$ and $I\sim24$, 
except the NF1 $R$ data which has a limit of $\sim24.5$.  This latter limit
is more comparable to the $V$ and $I$ limits for galaxies at
intermediate redshifts ($z\sim0.5-0.7$).  
To test the completeness levels of galaxy detection 
at these magnitude limits we checked the number of ``dark'' galaxies 
detected by FOCAS (or negative fluctuations on the CCD) in each field 
and bandpass.  The excess of above-sky objects 
to the number of below-sky objects 
detected can be used as an unbiased estimate of the number of faint objects at that 
magnitude (Valdes \markcite{V83}1983).  All of our samples 
were found to be essentially $\sim100\%$ complete at the magnitude 
limits that are listed above, confirming the conservative nature of  
our faintness limits.  
Photometric errors 
were calculated to be typically $\lesssim0.1$ mag. for galaxies 
with $V<24$, $R<24$ and $I<23$, increasing to as much as $\sim0.3$ mag. for  
fainter galaxies.  
In Table 5, the number of objects found 
within the given magnitude limits are tabulated, along with the effective 
field areas.  All of these objects 
comprise the magnitude limited samples which are used for angular
correlation analysis in \S 5.  

Color-selected samples were obtained by matching the catalogs generated
for a given field for the two bandpasses in the required color.  
The magnitude limit for each color-selected sample was chosen to be the 
limit of the bandpass (either $V\sim25$, $R\sim25$ or $I\sim24$, except 
NF1 where it's $R\sim24.5$) 
with the largest number of galaxies, thereby maximizing the accuracy 
of the clustering measurements.  This approach introduces some galaxies 
into the sample which are beyond the magnitude limit for the bandpass
with the smaller sample size, in order to maintain a high level of 
matching.  Colors are calculated using ``hybrid'' magnitudes 
but the number of galaxies measured with isophotal 
magnitudes is small enough such that there is little effect on our final results.  
This was checked by calculating the results using galaxy samples 
which contained only aperture 
magnitudes.  No attempts were made to convolve the two bandpass images to the 
poorest seeing before the colors were determined since the majority of 
galaxies easily fit into the $3''$ apertures adopted and the 
seeing values in different bandpasses for the same field are comparable 
(Table 2).  
High percentages (typically $\geq90\%$) of the galaxies matched between 
filters, particularly for the NF1 field where the 
magnitude limits in $V$, $R$ and $I$ were of similar depth.  A 
$5-10\%$ 
decrease in the match success rate was observed for the faintest
magnitude bins.  These faint galaxies are not detected in all three filters 
partly because of color effects and because the counts are beginning 
to become incomplete due to fluctuations in the background.  
Histograms of the colors of the galaxies found in the three
fields are shown in Fig. 7.  $(V-R)$, $(R-I)$ and $(V-I)$ galaxy colors
are shown in the plot panels from top to bottom for NF1 (solid line), 
NF2 (dashed line) and NF3 (dotted line).  
The shapes and modes of the 
histograms for a given color are very similar from field-to-field, 
with the differences
in the normalization due to the variations in the effective area of 
detection and in the galaxy numbers.  
The median colors of galaxies as a function of $R$ magnitude 
were found to be roughly consistent with those measured by 
Smail {et al.} \markcite{Sm95}(1995), 
in their Fig. 3.  These samples are used to 
measure the color-selected 
angular correlation function of galaxies.  

\section{Measuring the Angular Correlation Function}

\subsection{Estimator}

If one considers two differential elements of solid angle on the 
sky, $d\Omega_{1}$ and $d\Omega_{2}$, then the joint 
probability, $dP$, that galaxies 
will occupy the two elements with an angular separation of $\theta$ can be 
written as: 
\begin{equation}
dP=n^{2}(1+\omega(\theta))d\Omega_{1}d\Omega_{2},
\end{equation} 
where $\omega(\theta)$ is the angular correlation function 
and $n$ is the mean surface density of galaxies.  
A random (Poisson) distribution of galaxies 
yields $\omega(\theta)=0$ 
for all $\theta$.  
Therefore, $\omega(\theta)$ 
is simply a measure of the 
number of galaxy pairs observed at a given separation projected
on the sky, normalized by the number of galaxy pairs expected
if the galaxies are randomly distributed.  
The traditional estimator for $\omega(\theta)$ (Peebles \markcite{P80}1980), where
equal numbers of galaxies and random points are considered, 
is of the form:
\begin{equation}
\widehat{\omega}(\theta)=\frac{DD(\theta)}{RR(\theta)}-1
\end{equation}
where $DD(\theta)$ and $RR(\theta)$ 
represent the number of data-data and random-random pairs
at the angular separation $\theta$ (hereafter pair symbols will  
implicitly be assumed to be functions of $\theta$ and the 
hat symbol is used to denote an {\it estimate} of a function).  
Another estimator includes a cross-correlation of
the data and random objects:
\begin{equation}
\widehat{\omega}(\theta)=\frac{2DD}{DR}\frac{(N_{R})}{(N_{D}-1)}-1
\end{equation}
(see EBKTG\markcite{E91}, 
Infante \& Pritchet \markcite{IP95}1995, and references therein), where
$N_{D}$ and $N_{R}$ are the number of galaxies and random objects, respectively.
The estimators in equations (2) and (3) have 
greater than Poissonian variance, so, to minimize the 
noise we adopt the 
estimator suggested by Landy 
\& Szalay \markcite{LS93}(1993, hereafter LS; also see Hamilton \markcite{H93}1993): 
\begin{equation}
\widehat{\omega}(\theta)=\frac{(DD-2DR+RR)}{RR}.
\end{equation}  

For a given galaxy sample, 100 
files of random positions, each 
containing the same number of ``galaxies'' as those observed, 
were generated, yielding $\sim0.5-1.0~{\rm x}~10^{5}$ random objects.  
An increase in the number of random position files from 100 to 1000 
did not significantly improve the accuracy of the final $\omega(\theta)$
estimation but substantially increased the computing time.  
The random position files were created using the 
same detection mask which was used for the galaxy image (in all three bandpasses 
for a given field) 
to block out saturated objects and areas of the CCD with defects 
and vignetting.  
A mean of the 100 random files cross correlated with the particular galaxy sample
yielded DR in the estimator.  In addition, by averaging the files, one
can calculate the probabilities of obtaining a pair and triplet at 
a separation $\theta$ (LS\markcite{LS93} 
refers to these probabilities as $G_{p}(\theta)$ 
and $G_{t}(\theta)$, respectively), which are quantities required 
to calculate $\omega(\theta)$ and the errors associated with the estimator.  

To check the clustering estimator in equation (4), 
we also determined counts in cells using:
\begin{equation}
\widehat{\omega}(\theta)=\frac{<N_{i}N_{j}>}{<N_{i}><N_{j}>}-1,
\end{equation}
with the number of galaxies in cells i and j denoted by $N_{i}$
and $N_{j}$, and the angular brackets representing an average of 
all the cells with an angular separation within the bin 
$\theta\pm\delta\theta$.  Images were divided into square cells, 
$5''$ on a side.  
Excellent agreement was obtained 
between the data pairs (eq. [4]) and cell counts estimators, so, in 
the remainder we 
only refer to the data pairs approach.   

\subsection{The Clustering Model}

The standard model for the angular correlation function 
is: 
\begin{equation}
\omega(\theta)=A_{\omega}\theta^{-\delta},
\end{equation}
where $\delta$ has been found to range from $\sim0.6-0.8$ for faint samples.  
The value of $\delta$ could be 
dependent on the angular scales ($\theta$) probed 
or the magnitude limit of the galaxy sample 
(Maddox {\it et al.} \markcite{M90}1990, 
BTBJ\markcite{B94b} and Neuschaefer \& Windhorst \markcite{NW95}1995).  
If we want to use the estimated $\omega(\theta)$ projected on the 
sky to determine the {\it spatial} two-point correlation function, 
we must adopt a model which includes possible 
clustering evolution with redshift ($z$).  
The conventional model for the spatial 
correlation function is:
\begin{equation}
\xi(r,z)=\left(\frac{r}{r_{0}}\right)^{-\gamma}(1+z)^{-(3+\epsilon)},
\end{equation} 
where evolution with redshift is parameterized by $\epsilon$, 
$r$ is the proper length, and $r_{0}$ is the correlation length for $z=0$ 
(see Phillipps {\it et al.} \markcite{P78}1978, 
Peebles \markcite{P80}1980, Efstathiou {\it et al.} \markcite{E91}1991, 
Infante \& Pritchet \markcite{IP95}1995 and Hudon \markcite{H95}1995).  If the 
redshift distribution ($dN/dz$) is known for the magnitude limits
for which the angular correlation function has been measured, the relationship
between $\omega(\theta)$ and $\xi(r,z)$ is determined by an integral  
known as Limber's equation (see, for example, 
Peebles \markcite{P80}1980 and HL\markcite{HL96}): 
\begin{equation}
\omega(\theta)=Cr_{0}^{\gamma}\theta^{1-\gamma}\int_{0}^{\infty}D(z)^{1-\gamma}g(z)^{-1}(1+z)^{-(3+\epsilon)}\left(\frac{dN}{dz}\right)^{2}dz\left[\int_{0}^{\infty}\left(\frac{dN}{dz}\right)dz\right]^{-2}.
\end{equation}    
$D(z)$ is the angular diameter distance defined as:
\begin{equation}
D(z)=\frac{c}{H_{0}}\frac{q_{0}z+(q_{0}-1)(\sqrt{1+2q_{0}z}-1)}{q_{0}^{2}(1+z)^{2}},
\end{equation}  
with $g(z)$ and $C$ given by:
\begin{equation}
g(z)=\frac{c}{H_{0}}((1+z)^{2}(1+2q_{0}z)^{\frac{1}{2}})^{-1},
\end{equation}
and
\begin{equation}
C=\sqrt{\pi}\frac{\Gamma((\gamma-1)/2)}{\Gamma(\gamma/2)},
\end{equation} 
with $q_{0}$, $H_{0}$ and $\Gamma$ being the deceleration 
parameter, Hubble constant and Gamma function, respectively.  
Note that equation (7) and Limber's equation (eq. [8]) gives 
$\delta=\gamma-1$.  If $\gamma\simeq1.8$, 
the value derived from 
surveys of 
bright, nearby galaxies (Davis \& Peebles \markcite{DP83}1983), in the 
case of 
clustering which is fixed in comoving coordinates 
then $\epsilon=\gamma-3=-1.2$.  ``Stable clustering'', 
where 
the clustering is fixed in proper coordinates, is the result
when $\epsilon=0$.  If $\epsilon>0$, then there is 
a growth in the clustering with redshift in proper coordinates.  
Using an extrapolated redshift distribution ($dN/dz$) 
from the CFRS (Lilly {\it et al.} \markcite{L95a}1995a), 
with the same faint magnitude limit as our galaxy sample ($I\leq24$), 
we can solve 
Limber's equation for assumed values of $q_{0}$, $H_{0}$, $\gamma$ 
and $\epsilon$.  
The growth of clustering ($\epsilon$) can then 
be estimated by comparing the models with the observations of 
$\omega(\theta)$ (see \S 5.3).  

\subsection{Star Removal}

The number counts of objects at faint magnitudes are dominated by 
galaxies but at bright limits the stellar component makes
a significant contribution.  To correct for this stellar contamination
we plot a shape parameter vs. the aperture magnitudes
for all the detected objects.  This shape parameter 
is simply the difference between the ``core'' and 
aperture 
magnitude of an object, where the former
is the magnitude corresponding to the flux incident on the inner 3x3 pixels.  
The core and aperture magnitude difference 
is a measure of the object's light concentration 
and is analogous to Kron's \markcite{K80}(1980) $r_{-2}$ statistic 
which is proportional to the half-light radius.  

A plot of the core--aperture vs. aperture ($3''$) 
magnitudes for $V$, $R$ and $I$ in NF2 and $I$ in NF3, is given in 
Fig. 8 (see also Fig. 1 in 
Woods {\it et al.} \markcite{W95}1995, for NF1 in $I$).  
The stellar sequences 
are demarcated with a solid-lined rectangular box 
and at the brightest 
magnitudes they sometimes exhibit increasing values for the shape
parameter due to saturation.  Magnitude limits for each 
galaxy sample 
are shown by dashed vertical lines.  
Stars brighter than $V$, $R$ or $I=22$ are removed
from the final galaxy samples for all three fields, except in NF1
where stars are removed down to $R$ or $I=21$.    
With these 
limits the star/galaxy separation is unambiguous.  For fainter objects 
no attempt is made to further eliminate stars from the sample since 
compact galaxies could be mistakenly removed and 
stellar numbers
are very small relative to the galaxies at these faint limits.  
Objects identified as stars in one bandpass, for a given field, 
are removed from all the galaxy samples determined with the three filters.  
In NF1 and NF2, the vast majority of stars found in one filter are also identified 
in the remaining two filters.  
The stars in NF3 are identified 
only with the $I$ data because the $V$ and $R~$ PSFs were noticeably 
variable across the field.  This isn't 
a concern in view of the 
overlap between bandpasses for the stellar samples found  
in NF1 and NF2.  At brighter magnitudes where the stars are 
removed from the initial object lists, there is reasonable agreement 
between the Bahcall \& Soneira \markcite{BS80}(1980) model and the 
observed star counts in the three fields.  

\subsection{Integral Constraint}

A correction must be made for the integral constraint, which is  
due to the 
estimation of the density of galaxies, at a given magnitude limit, 
with a bounded, finite sample (Peebles \markcite{P80}1980).  
This bias has
the effect of reducing the amplitude of $\omega(\theta)$.  
Following BSM\markcite{BSM95} and LS\markcite{LS93} we calculate the integral
constraint ($\omega_{\Omega}$) using:
\begin{equation}
\omega_{\Omega}=\frac{1}{\Omega^{2}}\int\int\omega(\theta)d\Omega_{1}d\Omega_{2}
\end{equation}
with $\Omega$ representing the solid angle of the masked field.  We also
assume the angular correlation function has the functional form given
in equation (6).  
Integral constraints are calculated 
for each field and for power laws with $\delta$ ranging from $0.5-0.9$,
resulting in $\omega_{\Omega}\sim0.08A_{\omega}-0.01A_{\omega}$.  For
$\delta=0.8$, the integral constraints
are determined to be 
$\omega_{\Omega}\simeq0.0195A_{\omega},~0.0199A_{\omega},~0.0193A_{\omega}$
for NF1, NF2 and NF3, respectively.  
The values are comparable
since the field sizes and geometries are similar.  
This correction
is significant compared to the small amplitudes of $\omega(\theta)$ measured
from the faint galaxy samples (see \S 5).  

\subsection{Star Dilution Correction}

Since stars have not been removed from the photometric samples fainter than
$V$, $R$ or $I\sim22$,  
a correction must be made to the amplitude of $\omega(\theta)$ to account
for the stellar component present down to the magnitude limit.  
The number of stars expected at faint magnitude limits is 
taken from the 
model of Bahcall \& Soneira \markcite{BS80}(1980).  The three 
fields in this study were chosen to be at high Galactic latitudes where stars are
relatively scarce and therefore the stellar dilution corrections are small.  
The amplitude of $\omega(\theta)$ after the correction for stellar dilution
is made is given by:
\begin{equation}
A_{\omega}^{sc}=\left(\frac{N_{obj}}{N_{obj}-N_{s}}\right)^{2}A_{\omega}
\end{equation}
where $N_{obj}$ is the number of objects used to calculate $\omega(\theta)$,
$N_{s}$ is the number of stars predicted by the Bahcall \& Soneira 
model and $A_{\omega}$
is the ``raw'' amplitude of $\omega(\theta)$ before any corrections or 
weighting have been applied.  
Stellar dilution correction terms calculated for the various 
magnitude limited samples are listed in Table 7.  

\subsection{Combining Fields, Error Analysis and Fitting}

Since the use of just 
one of the three observed deep fields leads to a
determination of $\omega(\theta)$ 
which is of low accuracy, a 
strategy must be adopted to combine the data sets and to calculate
the average or ``final'' $\omega(\theta)$ and the appropriate errors.
We follow a similar approach to that of 
Neuschaefer {\it et al.} \markcite{N95}(1995, hereafter NRGCI), where the 
average $\omega(\theta)$ for a given bin ($\theta$) is calculated 
using:
\begin{equation}
{\widehat{\omega}}_{fin}(\theta)=\frac{\sum_{i} \eta_{i}\widehat{\omega}_{i}^{IC}(\theta)}{\sum_{i} \eta_{i}},
\end{equation}
where the summations extend over fields $i=1,2,3$, and 
where the $\eta_{i}$ are the weights
for each field and include the stellar dilution correction and 
the galaxy number densities.  
We use the number densities of objects as weights 
since the total areas of the three fields are slightly different.  
The $IC$ superscript on the $\omega(\theta)$ estimates
for each field denotes that corrections have been made for the integral constraint.  
Corrections for higher order correlations (e.g., three-point correlation function) 
are disregarded since they are 
negligible relative to the values of the ${\widehat{\omega}}_{i}^{IC}(\theta)$ 
calculated in 
the individual fields.  

Although the estimator for $\omega(\theta)$ 
which we use has been shown to give
Poissonian variance for uncorrelated data by LS\markcite{LS93}, 
it does not necessarily 
follow that it behaves this way for {\it correlated} data.  
This was first pointed out
by Bernstein \markcite{B94}(1994) who also   
emphasized that most authors do not properly 
account for the interdependence between the various bins 
when estimating the uncertainty in their results.  
We calculate errors using a scheme outlined
by Fisher {\it et al.} \markcite{F94}(1994) where the covariance matrix for 
a particular estimate of $\omega(\theta)$ is determined with 
bootstrap resampling (Barrow, Bhavsar \& Sonoda \markcite{B84}1984).  
Alternatively, Bernstein\markcite{B94} (1994) 
derives an approximate analytical expression
for the covariance matrix, but the model fitting procedures
in either study are essentially equivalent.  As in NRGCI\markcite{N95}, 
for a given field and magnitude-limited sample, resampled estimates of 
the 
${\widehat{\omega}}_{i}^{IC}(\theta)$ are calculated 
by applying the estimator of equation (4) to a resampled 
list of galaxies with the same number of objects as the original, real sample.  
The 
resampled list is generated by randomly selecting galaxies 
from the original list {\it with} replacement, such
that a galaxy can be chosen anywhere from zero up to several times.  
For each magnitude-limited sample, 50 bootstrap-resampled estimates of 
${\widehat{\omega}}_{fin}(\theta)$ are made by averaging  
resampled estimates of ${\widehat{\omega}}_{i}^{IC}(\theta)$ 
calculated for 
the three fields.  
The final bootstrap errors for the different angular separation bins 
are simply given by the variance of the 
resampled estimates of ${\widehat{\omega}}_{fin}(\theta)$.  
Finally, a covariance matrix is generated 
so that the power-law model
(eq. [6]) can be properly fit to the clustering observations 
for each magnitude-limited sample, following the technique described
by Fisher {\it et al.} \markcite{F94}(1994, Appendix A).  

Since the galaxy samples are fairly small (see Table 5), we choose
to fix the power-law exponent ($\delta$) and only let the  
amplitude ($A_{\omega}$) vary when fitting the model (eq. [6])  
to the data.   
The $\chi^{2}$
minimization is analytic with one linear parameter in the model.  
The power-law exponent, $\delta$, has been
measured to range from $\sim0.6$ to $\sim0.9$ 
(Neuschaefer \& Windhorst \markcite{NW95}1995, 
and references therein) 
for galaxies at faint magnitude limits.  Accordingly, each 
${\widehat{\omega}}(\theta)$ calculated for a magnitude-limited
sample is fit with power laws between 
$\delta=0.5$ to $\delta=0.9$, in $0.1$ increments.  The $\chi^{2}$ statistic
calculated for each fit gives an idea of what the most appropriate
value for $\delta$ is, although the measurement is not well constrained.  
Our data favour larger values of $\delta$ ($\delta\sim0.8-0.9$), 
so we fix 
$\delta=0.8$ to ease comparison with other studies.  
Table 6 illustrates the relative insensitivity of the final amplitudes 
to fits with power laws having 
different values for $\delta$, in this case for the faintest magnitude-limited
samples in $V$, $R$ and $I$.  The $\chi^{2}$ values, 
which are for {\it five} degrees of freedom, decrease as 
the power law approaches $\delta=0.8-0.9$ but for $\delta>1.0$ the errors in
the amplitude fit increase dramatically.  
For decreasing values of $\delta$ 
the integral constraint increases (\S 4.4) and this is the primary reason the 
$\chi^{2}$ statistic increases dramatically,  
as listed in Table 6. 
Given the limited statistics 
of our sample, fixing $\delta=0.8$ seems to be the best approach 
but it should be emphasized that
the errors given for the fitted amplitudes using this technique are 
probably underestimates.  
Further details of the results of the model fitting are discussed 
in \S 5.1 below.  

For each magnitude-limited sample in a particular field, 
${\widehat{\omega}}(\theta)$ is
calculated for angular separations 
ranging from $10-126''$, within 6 equally spaced logarithmic bins.  
The binning and angular separation range were carefully chosen to optimize
the measurement of $\omega(\theta)$ given the available imaging data.  
The upper limit for $\theta$ was chosen to be roughly one-third of 
the angular extent of the smallest field, thereby avoiding border 
effects.  A lower limit of $10''$ yielded error bars for the smallest
angular separation bin which were roughly comparable to those obtained
for bins with a larger $\theta$.  A discussion of close galaxy pairs 
with smaller angular separations ($\leq10''$) is given in 
Woods {\it et al.} \markcite{W95}(1995).   

\section{Angular Correlation Function Results}

\subsection{Magnitude-Limited Samples}

Measurements of the angular correlation function for the magnitude-limited
samples defined in \S 3.4, for $V$, $R$ and $I$, are presented in Figs. 9, 10 
and 11 respectively.  
The correlation 
amplitudes generally decrease over the small range of magnitudes probed, 
most obviously
with the $R$ data.  The solid lines in Figs. 9-11 
are {\it fits} of the model, $\omega(\theta)=A_{\omega}\theta^{-0.8}$, 
to the data and the errors are calculated using bootstrap resampling, 
as described in the previous section.  
Amplitudes, $A_{\omega}^{fin}$, 
measured from the fits to 
${\widehat{\omega}}_{fin}(\theta)$ 
for the various magnitude ranges 
are listed in Table 7, 
and are scaled for angular separations ($\theta$) 
given in arcseconds.  
The stellar dilution corrections used for the 
galaxy samples are also listed in Table 7.  
To demonstrate the scatter of the uncorrected estimates of  
$\omega(\theta)$ between the three fields, the ``raw'' angular 
correlation functions are plotted for the $I$-band data in Fig. 12.  
The circles, triangles and squares are data points measured 
from the NF1, NF2 and 
NF3 fields, respectively.  
NF1 data points have been offset $0.01$ dex to the left, with 
NF3 data moved the same amount to the right, to improve the clarity of this plot.  
These data have not yet been corrected for either 
the integral constraint or for dilution due to stars.  The improvement of the 
agreement of the $\omega(\theta)$ estimates in the three fields towards 
fainter magnitude ranges is typical for the three bandpasses and is 
due to the larger numbers of galaxies in the faint samples.  Note that the 
ordinate range 
of Fig. 12 is smaller than Fig. 11 to better illustrate the similarities 
between the three fields. 
In Figs. 9-11 it can be seen that the faintest sample for each bandpass appears
to have a slightly flatter distribution than the assumed $\theta^{-0.8}$ power law.  
Since there is good agreement between the clustering measurements made in the
three fields 
at faint limits this effect could be real and possibly connected with the 
flattening of the power law slope of $\omega(\theta)$ observed by 
Neuschaefer \& Windhorst \markcite{NW95}1995).  As discussed in the 
previous section, our galaxy samples are of insufficient size to accurately 
determine 
both the amplitude ($A_{\omega}$) of $\omega(\theta)$ and the 
power law slope ($\delta$) simultaneously.  
Assuming $\delta<0.8$ introduces a larger 
integral constraint correction for the data plotted in Figs. 9-11 and 
thereby increases the $\chi^{2}$ value for the fit.  The adoption of 
$\delta=0.8$ is to 
ease the comparison with other studies but it is still typically the best fit 
when all the corrections to the data are made.  However, it 
should be noted that there is some evidence for a decrease in the 
value of $\delta$ at fainter limits and this should be investigated with 
larger samples of faint galaxy photometry.  

The values of the angular correlation functions
measured in $V$, $R$ and $I$ at the separation of $\theta=1'$  
are plotted in Fig. 13.  
The $R$-band data gives the strongest clustering signal 
but the variations are roughly comparable to the errors.  
Lilly {\it et al.} \markcite{L95a}(1995a) 
have measured the redshift distribution
of faint galaxies selected to a magnitude limit of $I\simeq22$ and 
also make extrapolations for $N(z)$ to limits as faint as 
those obtained for the photometry in this work.  
With these redshift distributions and the $I$-band
angular correlation analysis by Lidman \& Peterson \markcite{LP96}(1996, 
hereafter LP) 
at brighter magnitude limits, we use our $\omega(\theta)$
measurements in $I$ to constrain model parameters 
for the {\it spatial} correlation function in \S 5.3.  
Finally, since there have been many recent correlation studies of 
faint galaxies in $R$-bandpasses, these provide a comparison
for the clustering detected with $R$-filter data in this study.  

In Fig. 14, the angular correlation function normalized to 
$\theta=1\arcdeg$, using the standard power law model with $\delta=0.8$, 
is plotted as a function of the
$R$ magnitude limit.  
The measurements by different
groups are denoted by the various symbols which are keyed to 
the authors'
initials and year of the particular paper.  
The $\omega(\theta)$
amplitudes determined in this work are given by the solid circles
for $R=20-24,~20-24.5,~20-25$.  Magnitude transformations for 
BSM\markcite{BSM95} were made using their assumption of $R\sim~r-0.55$.  
For Couch {\it et al.} \markcite{C93}(1993) and others 
the conversions given by Yoshii {\it et al.} \markcite{Y93}(1993)
and Roche {\it et al.} \markcite{R93}(1993) yield $R$ magnitudes 
from the original 
$VR$ and $r_{F}$ values.  The only observation plotted in 
Fig. 14 which was not taken with a red filter is that of 
MSFR\markcite{M95}.  
Since the MSFR\markcite{M95} result
is the $\omega(\theta)$ amplitude with the faintest magnitude limit
yet measured from the ground, it is interesting to include it for 
comparison using the approximate relationship 
$B_{CCD}\simeq~R+1$.  

It is notable that the $\omega(\theta)$ values 
presented in this work form a smooth continuation of the previous
observations made by Infante \& Pritchet \markcite{IP95}(1995) and HL\markcite{HL96}
for $R=21-23.5$, where the latter study used the same 
$R$ filter as the current observations.  
Our data agrees reasonably well with the 
Efstathiou {\it et al.} \markcite{E91}(1991)
data point and very well with the overlapping observations of 
Roche {\it et al.} \markcite{R96}(1996).  The largest discrepancy with this 
study is seen 
with BSM\markcite{BSM95}'s $\omega(\theta)$ 
measurements where our clustering amplitudes are 
observed to be factors of $\sim2-3$ larger.  
A possible explanation for part of this difference is that the 
BSM\markcite{BSM95}
field 
is at low Galactic latitude ($b\simeq35\arcdeg$) 
typically requiring larger stellar contamination
corrections than our three high Galactic latitude fields.  
Another possibility is that 
our clustering amplitude errors are underestimates 
since they are the fitting errors for a $\theta^{-0.8}$ power law.  
Nevertheless, the clustering measurements 
in this study are in agreement with 
Roche {\it et al.} \markcite{R96}(1996), 
who in turn agree with the BSM\markcite{BSM95}
results, so there is a reasonable level of consistency between studies 
in the $R\sim24-25$ magnitude range.  
With three fields 
the current work is less susceptible to variations
in general clustering behaviour induced by large scale structure.  

The fact that our $V$ and $I$-band estimates of $\omega(\theta)$ (Fig. 13) 
do not show well defined decreases with magnitude, as the $R$-filter data does, 
is not surprising due to the 
poorer statistics of the  
$V$ and $I$ filters for the two brightest magnitude bins (see Table 5).  
Therefore, the apparent flatness of the clustering 
amplitudes with magnitude 
for the $V$ and $I$ galaxy samples should not be interpreted 
as a strong trend but merely a clustering measure over  
a small range of magnitudes (note the galaxy samples 
are cumulative towards fainter magnitudes, not differential).  
Since 
clustering over a significant magnitude range cannot be tracked 
with just the data from this study, other studies must be included for 
a proper analysis of galaxy clustering evolution.  This 
is done for the $I$-band data in \S 5.3. 

Our Fig. 14 follows the format of Fig. 2 presented in 
BSM\markcite{BSM95}.   
Even with the additional results included from recent studies, there
still isn't general agreement on the 
precise slope of the monotonic decrease in clustering amplitude 
with limiting magnitude.  
No clear indication that the amplitude 
is starting to level off 
at faint magnitudes
is observed, as would be expected in some merger models of 
galaxy evolution (Carlberg \& Charlot \markcite{CC92}1992) or 
if there was a magnification bias
from weak gravitational lensing (Villumsen \markcite{V95}1995).  
Roche {\it et al.} \markcite{R93}(1993) and 
Roukema \& Yoshii \markcite{RY93}(1993, hereafter RY) have also 
found merging-model clustering behaviour to be inconsistent 
with current measurements of $\omega(\theta)$.  
MSFR\markcite{M95} 
claim that the amplitude of $\omega(\theta)$ flattens for $B$ data at about the same 
magnitude where the slope of the number counts flattens 
($B\sim25$ or $R\sim24$), which they attribute to an effective redshift
cutoff for the galaxies.  The data from various groups 
plotted in Fig. 14 shows  
that any flattening in the clustering amplitude with magnitude is not that well 
constrained as yet, especially considering 
the inherently large random and systematic 
errors which plague the measurement of $\omega(\theta)$ at faint limits.  

One cannot rule out the viability of {\it all} merger models of 
galaxy evolution with current observations but models with extreme amounts 
of merging or ``maximal merging'' as in RY\markcite{RY93} can 
be shown to be inconsistent with measurements of $\omega(\theta)$.  
(Note that merging is considered ``maximal'' with the assumption  
that every
galaxy merges when its dark halo merges.)  
See RY\markcite{RY93} for more details on the nature of their 
galaxy merging models.  
The clustering measurements for the 
increasingly faint magnitude-limited samples in the $V$ bandpass 
which are listed in Table 7 and shown in Fig. 13 
(with respective median magnitudes of $V_{median}=23.33,~23.84,~24.30$) 
have 
corresponding predictions from the RY\markcite{RY93} merger models 
which are 
${\rm log}(\omega(\theta=1',~\delta=0.8))\simeq-1.06,~-1.14$ and $-1.25$.   
Thus, a discrepancy of at least a factor of $\sim3$ is evident between 
our $V$-band $\omega(\theta)$ measurements 
plotted in Fig. 13 and the 
models plotted in RY\markcite{RY93}'s Fig. 3.  
Other observations which seem to rule out the most extreme 
merger galaxy evolution 
models (Carlberg \markcite{C95}1995) are a collection of redshifts for very faint 
galaxies obtained with the Keck telescope (Koo {\it et al.} \markcite{K96}1996).  
This spectroscopic sample is still sparse, so the results should be treated
as preliminary, but the median redshifts obtained for $I>22$ are contrary to 
what is expected for Carlberg\markcite{C95}'s ``maximal merging model''.    
The extent of the role of galaxy merging is still not 
clear but we can conclude that a {\it maximal} merger model is no 
longer a viable mechanism for galaxy evolution at intermediate redshifts.  

\subsection{Color-Selected Samples}

With the {\it multi-color} deep imaging in each field, 
the angular correlation function
can be measured with color-selected samples down to faint magnitude
limits.  A few approaches for obtaining color-selected samples were 
attempted to maximize the number of galaxies and thereby improve
the accuracy of the $\omega(\theta)$ measurements over a significant
range in color.  However, it should be noted that 
having just $V$, $R$ and $I$ images to work with, and 
no bluer bandpasses, unfortunately leads to a limited baseline of 
observed colors for the faint galaxies.  In Fig. 15, the
angular correlation functions selected by $(V-R)$, $(R-I)$ and $(V-I)$ 
colors are presented in a plot which is analogous to Fig. 1 of 
LSK\markcite{LSK96}.  
Amplitudes measured for the full range 
of angular separations, $10-126''$ (yielding an effective separation 
of $\sim35''$), are plotted as a function of the 
colors which the galaxies observed 
are either {\it less} than (top row of plots), or
{\it greater} than (bottom row).  
Poisson error bars are shown but should be considered to be underestimates of the
true errors.  The final amplitudes have been corrected for the integral constraints
and stellar dilution factors calculated for each field.  Also, the plotted values
are obtained with weighted averaging, where the weights are determined from 
the number of galaxies detected in both bandpasses for a given color.  

LSK\markcite{LSK96} used $(U-R_{F})$ colors 
to find that the $\omega(\theta)$ amplitude increased by over a factor
of ten for the reddest {\it and} bluest galaxies taken from a 
bright magnitude limited sample ($B_{J}\leq23.5$).  Fig. 15 shows
no indication of this behaviour for the current, fainter galaxy sample with
$(V-R)$, $(R-I)$ and $(V-I)$ colors.  Within the errors, the angular 
correlation amplitude integrated over the full range of separations 
is relatively constant regardless of the colors of the galaxies being
analysed.  This is consistent with what 
BSM\markcite{BSM95} ($R\lesssim25.5$) 
and Infante \& Pritchet \markcite{IP95}(1995, $b_{J}\leq24$ and $R_{F}\leq23$) 
have found although one should note that these previous studies made
only one division in color (blue/red) and did not look at the extremely
blue or red galaxies, as did LSK\markcite{LSK96}.  
It is possible the lack of an 
increase in the clustering amplitude 
in Fig. 15 is due to the $V$, $R$ and $I$ colors
not discriminating the reddest and bluest galaxies well enough 
with our coarser colour bins.  In other words,  
the numbers of galaxies available for this study may be simply
insufficient for providing an accurate measure of $\omega(\theta)$ with 
color.  There also 
could be physical reasons for the non-detection of a clustering
increase with extreme color, and these are discussed below.  

To do a more direct comparison of the clustering observed
with color-selected samples the 
correlation amplitudes for $(V-I)$-selected galaxy samples are
plotted in Fig. 16, for a fixed angular separation of $1'$.  
Results from NRGCI\markcite{N95} 
are given as open symbols for the 50\% blue, 50\% red and entire
samples while the filled symbols are measurements for 
objects in the current work with $(V-I)<1.3$, $(V-I)>1.3$ and for the full
sample, with the {\it median} $I$ magnitude plotted on the abscissa.  
The value $(V-I)=1.3$ was chosen as the dividing line such that the entire
sample could be cut into roughly 50\% blue and red galaxies.   
In this case, the $\omega(\theta)$ amplitudes calculated from our sample 
assume 
that $\delta=0.7$ in order to ease the comparison with 
NRGCI\markcite{N95} and our 
error bars are calculated
using bootstrap resampling.  
Galaxies with the $50\%$ reddest $(V-I)$ colors in 
NRGCI\markcite{N95} were observed to 
have clustering which 
is $\sim4-8$ times stronger than the blue half of the sample, but 
with 
substantial errors.  
NRGCI\markcite{N95} also argue that
there is an increase in amplitude for the 20\% color marginals 
(bluest and reddest 1/5 of the galaxies) 
from the 50\% samples, although these two samples for either blue
or red objects are consistent within the errors.  
The $(V-I)$-selected sample in this study, 
which is one magnitude deeper than that of NRGCI\markcite{N95}, shows no sign of
color segregation of the clustering amplitudes beyond $I\sim22$, albeit
with large error bars for the measurements.  Also, there is no significant
difference observed between the clustering of the galaxies in the 
full, $I$-selected sample and the blue and red samples.  The amplitudes 
from our red and blue galaxy samples do not bracket those calculated 
for the full 
sample due to not all the galaxies being detected in both the $V$ and $I$ 
images.  
Clustering amplitudes for 
the red and blue 20\% marginals in our sample were not determined since 
large errors would result from the small sample size.  

LP\markcite{LP96} also determined $\omega(\theta)$ 
for $(V-I)$-selected samples of galaxies, in the magnitude range $I=18-20$.   
They used $(V-I)=1.5$ as the blue/red boundary and 
found a marginally significant 
difference between the samples with red galaxies 
exhibiting stronger clustering, similar to the NRGCI\markcite{N95} 
results at brighter
magnitudes.  This comparison of LP\markcite{LP96}, 
NRGCI\markcite{N95} and the current work suggests
that a difference in the clustering amplitudes for blue and red galaxies 
(using $(V-I)$ selection) exists at bright magnitudes ($I\sim18-21.5$) and either
disappears or has not been detected at fainter magnitude limits ($I\sim21.5-24$).  
Interpreting these color-selected clustering results is 
complicated in that, for $I$-selected samples (for e.g., the CFRS with $I\leq22$), 
red galaxies tend to be confined to a fairly 
narrow range of intermediate redshifts 
while blue galaxies are observed to have more broadly distributed 
redshifts ($z\sim0-1$), 
with a lower mean $z$.  The results summarized in Fig. 16 may be showing  
that significant galaxy evolution is occurring at faint magnitudes 
relative to brighter magnitudes (lower redshift).  
Red and blue galaxies are observed to cluster differently at lower $z$, which 
is simply a reflection of the 
morphology-density relation.  Another possibility is that the blue sample
is more diluted with lower luminosity galaxies which have stronger clustering 
properties, making the clustering measurements for the blue and red faint galaxy 
samples indistinguishable.  A more accurate approach for tracing clustering 
evolution of ``typical'' $L^{*}$ galaxies may be to select out red galaxies
and measure their clustering variations with magnitude, since the 
luminosity function of these objects shows very little change over $0<z<1$ 
(Lilly {\it et al.} \markcite{L95a}1995a).  
Obviously, larger multi-color imaging surveys of  
faint galaxies are required to more accurately
measure the color-selected angular correlation function and further check the
viability of various galaxy evolution scenarios.  

\subsection{Comparison with Models of the Spatial Correlation Function}

To determine a viable model for the evolution of $\xi(r)$ with redshift (eq. [7]) 
from measurements of $\omega(\theta)$, one requires
the redshift distribution ($dN/dz$) of the galaxies 
within the magnitude interval 
being considered.  With a realistic redshift distribution, Limber's 
equation (8) 
can be solved and relationships between the observed $\omega(\theta)$ 
and inferred $\xi(r)$ can be determined for different cosmologies ($q_{0},H_{0}$) and 
clustering evolution ($\epsilon$), with the power law index for $\xi(r)$
constrained from the angular clustering results ($\gamma=\delta+1$).  
Unless otherwise noted we assume that $q_{0}=0.5$ and 
$H_{0}=100~{\rm km~s}^{-1}~{\rm Mpc}^{-1}$.  
Since the correlation length at $z=0$, $r_{0}$, 
corresponds to the amplitude of the locally observed $\xi(r)$, 
a range of reasonable values for $r_{0}$ and
$\epsilon$ are assumed in order to generate models of the clustering evolution.   
These models are then compared to 
the observed values of $\omega(\theta)$.  

The $I$ photometric data in this study have a magnitude limit ($I\sim24$)
which is a full
two magnitudes fainter than the currently largest deep redshift survey
(Lilly {\it et al.} \markcite{L95b}1995b).  For an estimate of 
the redshift distribution at the limits of the photometry, the $dN/dz$ 
measured to $I\sim22$ can be extrapolated to fainter magnitude limits 
with a no-evolution assumption for the galaxies.  
Evolution is obviously occurring for the galaxies at some level  
towards fainter magnitudes but the discrepancy between the observed
galaxy counts and extrapolated-$dN/dz$ 
number counts is fairly small for the 
two magnitude interval beyond $I\sim22$ (see Figs. 8 and 9 in 
Lilly {\it et al.} \markcite{L95a}1995a).  
Using the Lilly {\it et al.}\markcite{L95a} extrapolations to $I\sim24$
and the observed redshift distributions for brighter magnitude limits, 
we have calculated 
the variation of $\omega(\theta)$ with $I$ magnitude for given values
of $r_{0}$, $\epsilon$ and $\gamma$.  

Errors are almost certainly present in the extrapolated redshift distributions
used to calculate the clustering evolution models.  Since 
the amplitude of $\omega(\theta)$ calculated using Limber's equation 
(eq. [8]) has a strong dependence on the shape (essentially the width) of 
$dN/dz$, errors will occur if this shape is 
poorly estimated with the extrapolation, while the effect of 
an inaccurate normalization will be small.  
Preliminary observational support for the extrapolated 
Lilly {\it et al.} \markcite{L95a}(1995a) 
redshift distributions has been provided by the DEEP survey 
(Koo {\it et al.} \markcite{K96}1996) where the median redshifts {\it measured} 
at faint limits are found to be consistent, 
albeit within large 
error bounds.  
Hudon \markcite{H95}(1995) and HL\markcite{HL96} 
have shown that if the redshift distribution, with slightly brighter magnitude 
limits than ours, 
has $15\%$ more galaxies added to it which are similarly distributed in redshift
there is little change in 
the estimate of the spatial correlation function, as expected.  
In the more extreme 
scenario where  
this $15\%$ is added at median redshifts of 
$z_{med}=1.5$ or $2.1$ as a gaussian distribution the resulting 
correlation length $r_{0}$ is increased by $\sim15\%$ or $\sim30\%$, respectively.  
This is probably a reasonable upper bound for the uncertainty in this study due 
to our still sparse knowledge of the redshift distribution of galaxies 
with $22<I<24$.  
Also, the shape of $dN/dz$ 
at these faint magnitudes will be incorrect if a 
particular galaxy population dominates at these limits but is not detected
at brighter magnitudes.  The calculations using Limber's equation are  
presented keeping these caveats in mind.  

LP\markcite{LP96} 
have measured $\omega(\theta)$ for a wide
range of brighter magnitudes in $I$ and we use these results as a 
comparison to this study, as well as to the models.  
In Fig. 17 the logarithm of $\omega(\theta)$ at an
angular separation of $1'$ is plotted with the {\it median} $I$ magnitude
given on the abscissa.  All of the samples obtained from the 
two fields (CL and FBS) in LP\markcite{LP96} are included, in addition to the 
Efstathiou {\it et al.} \markcite{E91}(1991) point for the $I$-band.  
Models calculated using the aforementioned redshift distributions 
from Lilly {\it et al.} \markcite{L95a}(1995a) are plotted as a series of lines 
for different values of $r_{0}(z=0)$ ($5.4~h^{-1}~{\rm Mpc}$ solid lines, 
$4~h^{-1}~{\rm Mpc}$ dashed lines and $2~h^{-1}~{\rm Mpc}$ dotted lines) and 
$\epsilon$ ($-1.2,0,1,2$ from top to bottom 
for each set of lines with a given $r_{0}$).  
The value of $\gamma$ is fixed at 1.8 following the discussions in
\S\S 4.2 and 4.6, but the effects of varying it 
are shown below.  
The LP\markcite{LP96} results are amplitudes obtained from galaxies within
narrow luminosity bins (1 or 2 magnitudes wide) spanning $I=16-23$ while 
the points determined in the current study are for galaxies with
$I=19-23,23.5,24$.  
Our magnitude limits were chosen to minimize the
error in the $\omega(\theta)$ measurements since there are a limited
number of galaxies available (2697 with $I=19-24$).  
As noted earlier, the majority of the galaxies in our 
$I$-selected samples are within a narrow magnitude range so no 
claim is made 
for a detection of flattening in $\omega(\theta)$.  
When combined with the LP\markcite{LP96} clustering study, a generally smooth
decline in the amplitude of $\omega(\theta)$ with $I$ magnitude is observed.  
For the slightly better statistics of the $R$ observations (Fig. 14), a decrease 
in the clustering amplitudes with 
magnitude limit is unambiguously seen to the faintest limits, 
in conjunction with other studies.  
There is fairly good agreement between the correlation function
amplitudes from the three $I$-band 
studies at the faintest magnitudes in Fig. 17.  
At $I_{med}\sim22$ the LP\markcite{LP96} data are consistent with our measurement of 
the clustering.  For the LP\markcite{LP96} point at $I_{med}\sim22.5$ and the 
Efstathiou {\it et al.}\markcite{E91} result at $I_{med}\sim23$ there is agreement 
within $2\sigma$ of the amplitudes obtained from this study  
but with a substantial
error for the 
LP\markcite{LP96} measurement at their magnitude limit.  

A comparison of 
the observations to the models in Fig. 17 leads  
to some general conclusions.  Clustering evolution which
is fixed in co-moving coordinates ($\epsilon=-1.2$) is a 
viable scenario only if $r_{0}(z=0)\sim2-3~h^{-1}~{\rm Mpc}$.  Values 
for $r_{0}$ are typically not observed to be this small
for the entire galaxy population. The correlation length usually ranges 
from $\sim4~h^{-1}~{\rm Mpc}$, calculated using 
IRAS-selected redshift surveys
(Saunders {\it et al.} \markcite{S92}1992, 
Fisher {\it et al.} \markcite{F94}1994), to the 
canonical optical survey correlation length of 
$r_{0}(z=0)=5.4~h^{-1}~{\rm Mpc}$ from Davis \& Peebles \markcite{DP83}(1983).  
BSM\markcite{BSM95} 
show that, with a correlation length of $r_{0}\sim2~h^{-1}~{\rm Mpc}$, 
and a rate of clustering growth predicted by linear theory  
($\epsilon\sim0.8$), they can match their clustering observations and models 
at faint limits.  From this result, BSM\markcite{BSM95} claim that low 
surface brightness (LSB) and/or dwarf galaxies
are dominating the faint galaxy population since some local measurements 
of the correlation lengths for these objects yield 
$r_{0}\sim2.3-2.7~h^{-1}~{\rm Mpc}$ 
(Santiago \& da Costa \markcite{S90}1990).  
However, the values for the LSB/dwarf galaxy local correlation lengths are
still controversial and may be larger 
(Thuan {\it et al.} \markcite{T91}1991).  
As noted earlier (\S 5.1), 
the amplitudes of the 
BSM\markcite{BSM95} 
observations in the $R$-band are significantly lower than what 
is observed in this study at similar magnitude limits.  Given that most 
studies to date have found local correlation lengths with 
$r_{0}\gtrsim4~h^{-1}~{\rm Mpc}$ along with the assumption that 
faint galaxy populations 
evolve into locally observed galaxies, 
our $I$-filter observations then suggest that 
$\epsilon\geq0$, in agreement with 
HL\markcite{HL96}, 
Le F\`{e}vre {\it et al.} \markcite{Le96}(1996) and 
Shepherd {\it et al.} \markcite{S97}(1997).  

For a non-negative value of $\epsilon$, 
two general possibilities remain for the 
evolution of the faint galaxy
population.  
The first scenario is that 
$\epsilon\simeq 0-1$ and 
$r_{0}(z=0)\sim4~h^{-1}~{\rm Mpc}$, where 
the excess of faint blue galaxies 
is due to objects which are analogous to 
IRAS-selected galaxies with respect to star formation, morphology and clustering, 
as suggested by BTBJ\markcite{B94b}.  The second 
possibility is that $\epsilon>1$, implying significant evolution 
in the clustering from faint limits to locally observed galaxies 
such that 
a value of $\sim5-6~h^{-1}~{\rm Mpc}$ is found for $r_{0}$.   
This correlation length is in agreement with 
most optically-selected, local redshift survey measurements of $\xi(r)$.  
We note that Efstathiou {\it et al.} \markcite{E91}(1991) 
only considered clustering models 
with $-1.2\lesssim\epsilon\lesssim0$ in order 
to obey the standard gravitational
instability picture.  
However, 
more recent N-body studies (Melott \markcite{M92}1992, 
Yoshii, Peterson \& Takahara \markcite{Y93}1993)
have found that models with 
$\epsilon\sim0-3$ are indeed possible due to the continual merging of 
groups as the universe expands.  

To illustrate the sensitivity of the models for $\omega(\theta)$ in $I$ to 
the assumed $q_{0}$, $\gamma$ and $\epsilon$, for a given $r_{0}$, the 
observed clustering amplitudes are plotted again in Fig. 18 versus the $I$ magnitude 
limit along with three different families of models.  
Note that each ordinate of the three panels covers a different range 
of clustering amplitudes but has a total range of 1 dex.  
For each 
$r_{0}$ listed ($5.4,~4,~2~h^{-1}~{\rm Mpc}$ from top to bottom) 
in the lower left corner, 
the solid line corresponds to the model calculated for $q_{0}=0.5$, $\gamma=1.8$
and $\epsilon=0$.  Assuming a small-$\Omega_{0}$ universe with $q_{0}=0.1$ 
yields the dotted line model for each $r_{0}$.  Changing just the power-law
index for the correlation function to the two extremes of what is observed, 
$\gamma=1.9$ and $\gamma=1.6$, gives the short-long dashed lines above and 
below the solid line, respectively.  Finally, the long dashed lines are 
associated with changing just the value of $\epsilon$, as in Fig. 17.   
The long-dashed line above the solid line is for $\epsilon=-1.2$ while
the two dashed lines below correspond to $\epsilon=1,2$ for decreasing
amplitude.  Clearly the comparison between the observations and models
does not have a strong dependence on the assumed value of $q_{0}$.  
The clustering evolution 
parameter $\epsilon$  
provides the greatest leverage in parameter space 
for matching the observations to clustering models, 
in addition to being poorly constrained.  If the inherent 
degeneracy of fitting clustering models to measurements 
of $\omega(\theta)$ is to be broken, better determinations 
of $r_{0}$ and $\gamma$ for local samples of galaxies 
selected by morphology, luminosity and surface brightness are 
required.  

\section{Discussion}

Lilly \markcite{L93}(1993) summarized the three standard scenarios which were  
suggested to explain the obvious evolution in the galaxy number counts 
(Tyson \markcite{T88}1988, Lilly {\it et al.} \markcite{LCG91}(1991)) and the 
lack thereof in the measured redshift distributions 
(Broadhurst {\it et al.} \markcite{B88}1988, Colless {\it et al.} 
\markcite{Col90}1990, \markcite{Col93}1993).  
First, the faint 
blue galaxies are explained to be proto-dwarf galaxies undergoing 
bursts of star formation at intermediate redshifts ($z\sim0.4$) and then
evolving into galaxies at the faint end of the luminosity function 
($L<0.01L^{*}$) 
by the current epoch (Broadhurst {\it et al.} \markcite{B88}1988).  
The second model has faint galaxies 
being very shortlived, star-bursting objects 
which are subsequently disrupted or fade away in such a fashion 
that they aren't observed in large numbers at small redshift (Babul 
\& Rees \markcite{BR92}1992).  
The final conventional model invokes 
merging of sub-galactic units at intermediate redshifts where current 
$L^{*}$ galaxies are the products of this process (Broadhurst 
{\it et al.} \markcite{B92}1992).  
We refer to these three frameworks as the 
``bursting dwarfs'', ``fading dwarfs'' and ``merger'' models.  
By looking for increases in the clustering 
of faint galaxies using various selection criteria (e.g., magnitude-limited 
samples, small angular
separations, colors), 
we can test the 
viability of the merger galaxy evolution model, as in \S 5.1.  
Dwarf models can be checked 
by comparing the clustering behaviour of the various local and faint 
galaxy populations (\S\S 5.2 and 5.3).  
A more unconventional explanation for the 
number counts and 
redshift distributions evolution discrepancy is the proposal 
by McGaugh (\markcite{M94}1994, also see 
Ferguson \& McGaugh \markcite{FM95}1995) 
of the existence of 
a significant population of low surface brightness galaxies which are 
not typically detected locally, due to observational selection effects, but 
which can be detected in faint, photometric surveys.  

As shown in the previous section, 
neither a sustained flattening in 
the amplitude of the 
angular correlation function over a significant range of magnitude, nor 
a well defined change in the slope of the decreasing $\omega(\theta)$ with 
magnitude, is seen with current data, 
which suggests that merger models  
(Guiderdoni \& Rocca-Volmerange \markcite{GR90}1990, 
Broadhurst {\it et al.} \markcite{B92}1992, 
Carlberg \& Charlot \markcite{CC92}1992) 
may not be viable descriptions of faint galaxy evolution.  
It is still possible 
to incorporate some merging into galaxy evolution models without 
being inconsistent with the observed clustering and redshift distributions, 
but models with ``maximal merging'' are certainly ruled out 
(\S 5.1 and Koo {\it et al.} \markcite{K96}1996).  
A lack of a significant amount of merging of galaxies at faint limits is 
in agreement with our results from Woods {\it et al.} \markcite{W95}(1995),  
where no substantial excess of 
close pairs of galaxies in NF1 were found down to $I\leq24$.  
Obviously the clustering measurements displayed in Fig. 14 are still too 
inaccurate to reasonably constrain any detailed galaxy evolution model.  

Assessing whether the faint galaxy population are predominantly bursting dwarfs 
or fading dwarfs is very difficult (if not impossible) 
to distinguish at this juncture using only 
estimates of the clustering.  
These two scenarios can be respectively described as follows 
(see Fig. 17 and Table 8): 
(1) the majority of faint galaxies are evolving into 
a local population which is similar to IRAS-selected galaxies 
($r_{0}\sim4~h^{-1}~{\rm Mpc}$) 
with moderate clustering evolution ($\epsilon\sim0-1$) and 
(2) the clustering 
is fixed in comoving coordinates with $\epsilon\sim-1.2$ 
requiring the 
local counterparts of faint galaxies to be either 
low surface 
brightness galaxies which are 
weakly clustered ($r_{0}\sim2-3~h^{-1}~{\rm Mpc}$) or galaxies which 
have been disrupted and are therefore undetectable.  
Additional local surveys for low surface brightness galaxies are required 
to better constrain the purported weak 
clustering behaviour of this population. 
The third model, previously discussed, 
where merging of galaxies plays some role 
is given by: 
(3) clustering of the faint population 
evolving at a significant rate ($\epsilon>1$) yielding local galaxies with 
similar clustering properties to what is observed in optical surveys 
($r_{0}\sim5-6~h^{-1}~{\rm Mpc}$).  
The three approximate pairs of $r_{0}$ and $\epsilon$ which match 
the clustering model to our $I$-band $\omega(\theta)$ measurements, are 
also in broad agreement with those found by 
HL\markcite{HL96} in the
$R$ bandpass.  
We summarize these three galaxy evolution scenarios in Table 8 and 
comment on the models' strengths and drawbacks.  

Clustering evolution 
processes (1) and (3)  
could both be occurring 
but recent HST observations 
suggest that late-type and irregular galaxies dominate the number counts
at faint limits (Driver {\it et al.} \markcite{D95}1995, 
Glazebrook {\it et al.} \markcite{G95}1995).  
Loveday {\it et al.} \markcite{L95}(1995) 
have measured the local correlation lengths for early and late-type
galaxy morphologies to be 
$5.9\pm0.7~h^{-1}~{\rm Mpc}$ (E and S0) and 
$4.4\pm0.1~h^{-1}~{\rm Mpc}$ (Sp and Irr), respectively.  If spirals and 
irregulars are the dominant population comprising the faint counts it is 
possible that the primary path for faint galaxy 
evolution is scenario (1) above, with a moderate change in 
the clustering ($\epsilon\sim0-1$).  This value of $\epsilon$ is 
consistent with what is predicted by linear theory and is easier to account 
for than larger values of $\epsilon$.  

Another point which should be 
made is that some of the large fraction of irregular galaxies found at 
faint magnitude limits with HST data (Driver \markcite{D95}{\it et al.} 1995, 
Glazebrook {\it et al.} \markcite{G95}1995)  
could be galaxies at high redshift with $2.3\leq~z~\leq3.5$ 
(Abraham {\it et al.} \markcite{A96}1996), where the lower redshift limit 
corresponds to the Lyman discontinuity entering the $U$ filter.  This 
possibility is 
supported by 19 of the 83 ``irregular/peculiar/merger'' objects found by 
Abraham {\it et al.}\markcite{A96} in the Hubble Deep Field to have 
$(U-B)>-0.2$.  A red 
UV-optical, $(U_{{\tiny n}}-{\cal G})$, color 
criterion has also been recently 
used by Steidel {\it et al.} \markcite{St96}(1996) 
to find ``normal'', star-forming 
galaxies (with ${\Re}\leq25$) at redshifts of $3\leq~z~\leq3.5$, confirmed with
Keck spectroscopy.  
However, Steidel {\it et al.}\markcite{St96} have found the surface density 
of the galaxies in this redshift range 
to be quite low ($0.4\pm0.07~\rm{galaxies}~\rm{arcmin^{-2}}$).  This suggests that if 
the Abraham {\it et al.}\markcite{A96} irregular objects are at high 
redshift then the bulk of these galaxies will have redshifts 
with $2.3\leq~z~\leq3$.  
Galaxy morphologies 
at large redshifts are subject to K-corrections, evolutionary effects of 
stellar populations and a strong dependence on surface brightness 
(Giavalisco {\it et al.} \markcite{G96}1996).  
Considering these effects, 
it is not clear whether the significant fraction of faint galaxies 
with apparent irregular/peculiar morphology are {\it all} 
the result of dynamical evolution 
at intermediate redshifts or 
the galaxies are at high redshift ($z>2.3$) and are observed to be irregular 
due to these ``morphological K-corrections''.  
Morphology 
studies of local galaxies at UV wavelengths and further faint galaxy 
spectroscopy (e.g., Koo {\it et al.} \markcite{K96}1996) 
should address this concern.  

No evidence was found in this study for a dependence of clustering 
on galaxy color (Figs. 15 and 16), such as what 
LSK\markcite{LSK96} found using $U-R_{F}$ colors 
at brighter magnitude limits.  
Nevertheless, Roche {\it et al.} \markcite{R96}(1996) find a 
difference in the clustering of red and blue galaxies 
at $B\sim25.5$.  They suggest that the observed clustering is
consistent with the LSK\markcite{LSK96} 
result at $B\leq23.5$.  No significant differences
in the clustering measured for red and blue samples of galaxies were 
found by Efstathiou {\it et al.} \markcite{E91}(1991) and 
BSM\markcite{BSM95}.  NRGCI\markcite{N95}, 
as mentioned earlier, find only marginal differences in the clustering 
for $(V-I)$-selected 
samples of 
red and blue galaxies at faint limits ($I>21.5$) while larger discrepancies
are observed for brighter magnitudes ($I<21.5$).  This roughly consistent clustering 
of 
red and blue galaxies at faint magnitudes is in agreement with what is observed 
for the $(V-I)$-selected samples of this work (Fig. 16).  
It remains to be seen if the difference (or consistency) of the clustering between 
red and blue galaxies is a function of the sample magnitude limit.  
The accuracy of the measurements of the color-selected $\omega(\theta)$ in
the current work does not warrant further analysis.  
Larger photometric surveys of faint galaxies will be 
essential towards achieving 
more precise determinations of $\omega(\theta)$ for magnitude limited
and color-selected 
galaxy samples.  

\section{Summary}

The amplitude of 
$\omega(\theta)$ is found to decrease with magnitude limit 
when our $R$ and $I$-band 
results are combined with clustering determined by 
other authors.  
This observed monotonic decrease with magnitude 
rules out ``maximal merger'' galaxy evolution models.  
Angular correlation function estimates (in $I$) in the current 
study and LP\markcite{LP96} were compared to galaxy clustering evolution models, 
generated with CFRS redshift distributions which were extrapolated to the 
faint limits of the photometry.  The observed clustering of the 
faint galaxies can be explained with local correlation lengths for 
this population of $\sim4~h^{-1}~{\rm Mpc}$ or $\sim5-6~h^{-1}~{\rm Mpc}$ 
for moderate ($\epsilon\sim0-1$) or strong ($\epsilon>1$) clustering 
evolution, respectively.  Clustering evolution which is fixed in co-moving 
coordinates ($\epsilon=-1.2$) is possible but requires 
smaller correlation lengths ($\sim2-3~h^{-1}~{\rm Mpc}$) than what is 
usually observed for local galaxies.  No evidence is found for 
variations in clustering which are dependent on galaxy color in this
study.  Larger photometric surveys are required to confirm the 
stronger clustering amplitudes for faint, red galaxies found by 
Roche {\it et al.} \markcite{R96}(1996) and to improve upon the 
accuracy of angular correlation function measurements 
at faint magnitude limits.  

\acknowledgments

We thank the CFHT TAC for generous allocations of observing time.  
Kudos to Harvey Richer for helping collect the NF1 data and for contributing 
part of the initial (and ongoing) impetus for the high Galactic latitude fields 
project.  Simon Lilly and Dan Hudon are thanked for providing some of their 
images so that NF2 could be flat-fielded, and for sending the CFRS 
extrapolated redshift distributions in machine readable form.  
David Schade is thanked for promptly transferring the Lilly/Hudon data
when it was needed.  DW thanks Bob Abraham, 
Brad Gibson, Dan Hudon, Ted Kennelly and 
Frank Valdes for stimulating and helpful conversations.  
Finally, we would like to thank the referee, Ian Smail, for a 
detailed report which improved the clarity of this paper.  
Support for this research was provided by grants to GGF and 
Harvey Richer from the National Science and Engineering Research 
Council of Canada.

\clearpage

\figcaption[nf1.eps]{$V+R+I$ image of the NF1 field 
with each bandpass normalized
to an equivalent sky level. The areas of the CCD frame with 
cosmetic defects, saturated
stars and very bright galaxies are masked out and not considered in the image
analysis. 
Note the cosmetic differences between
the LICK1 CCD (NF1) and the LICK2 device (NF2 and NF3; Figs. 2 and 3).  
\label{fig1}}

\figcaption[nf2.eps]{$V+R+I$ image of the NF2 field with each bandpass normalized
to an equivalent sky level, as in Fig. 1. \label{fig2}}

\figcaption[nf3.eps]{$V+R+I$ image of the NF3 field with each bandpass normalized
to an equivalent sky level, as in Fig. 1. \label{fig3}}

\figcaption[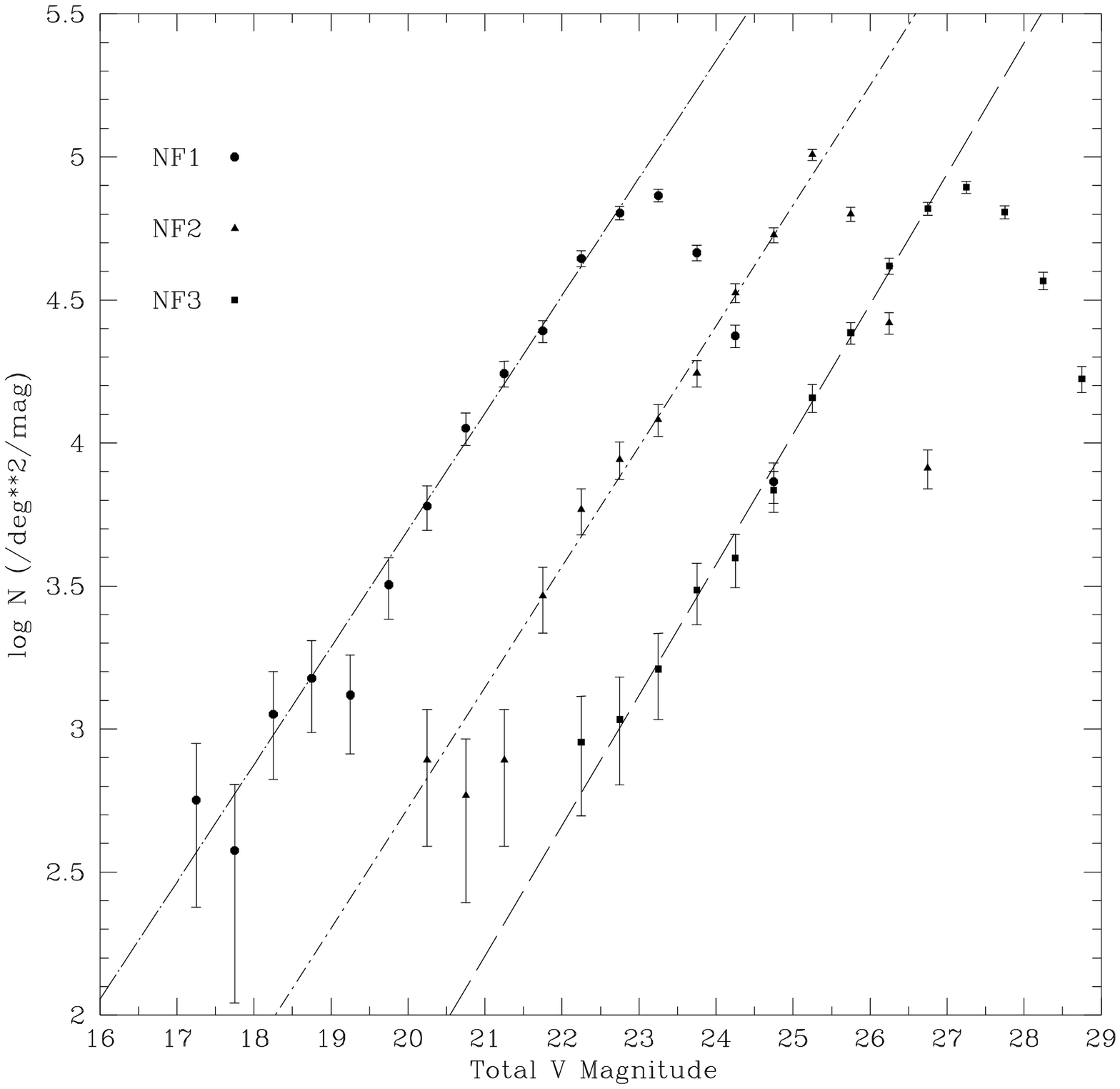]{$V$ number counts for NF1, NF2 and NF3. 
The plotted lines are least squares fits to all the points above the 
$V\sim25$ magnitude limits.  
Slopes are listed in Table 4.  NF3 counts have been shifted to the right and NF1 counts
to the left, by 2
magnitudes, for purposes of clarity.  NF1, NF2 and NF3 counts are denoted by
circular, triangular and square points, respectively.  The coordinate on the
abscissa corresponds to ``hybrid'' magnitudes, as discussed in the text. \label{fig4}}

\figcaption[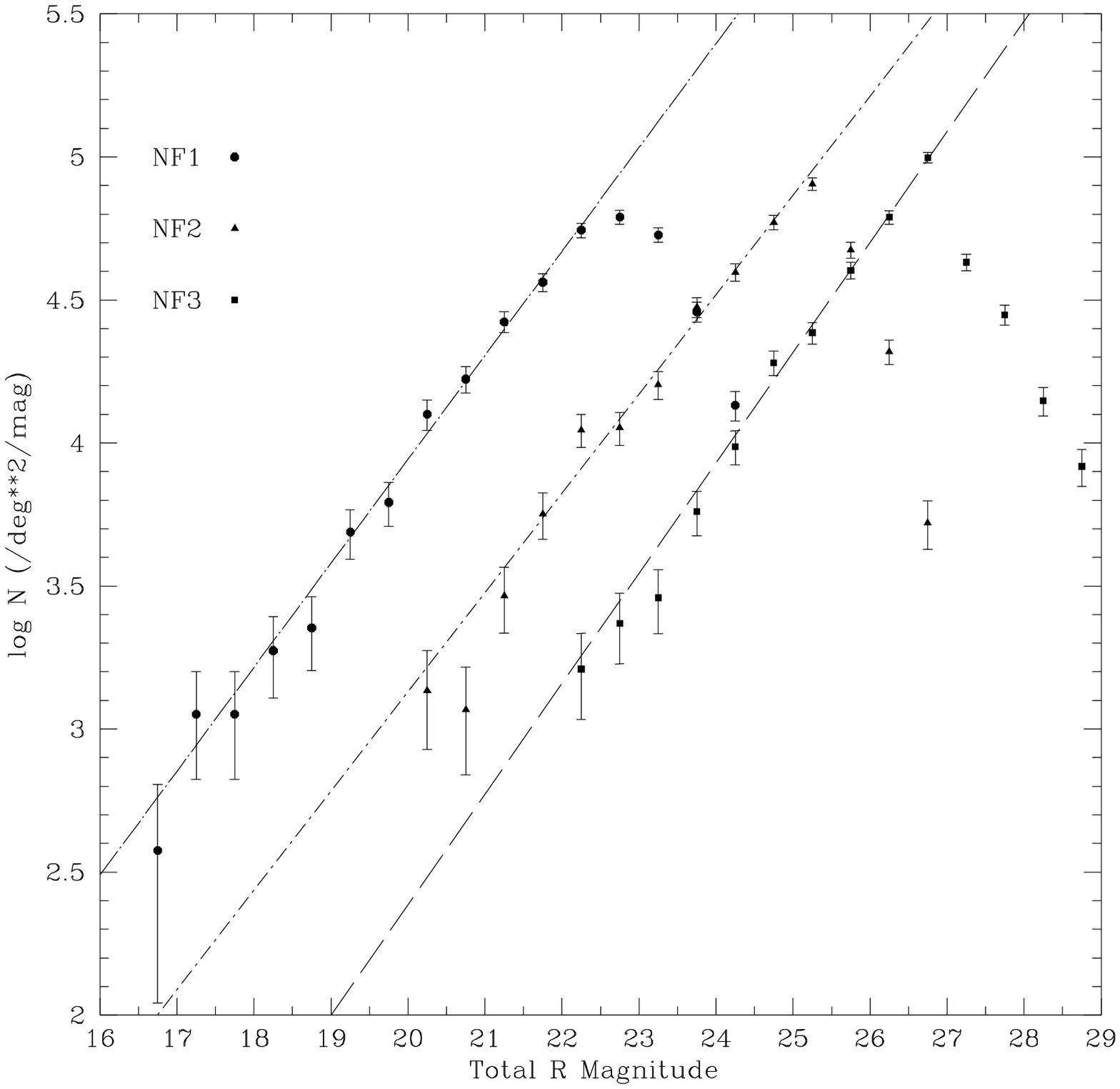]{$R$ number counts for NF1, NF2 and NF3. 
The plotted lines are least squares fits to all the points above the 
$R\sim25$ magnitude limits, except for NF1 which has a magnitude limit
of $R\sim24.5$.  Slopes are listed in Table 4 and the format of the plot
is the same as in Fig. 4. \label{fig5}}

\figcaption[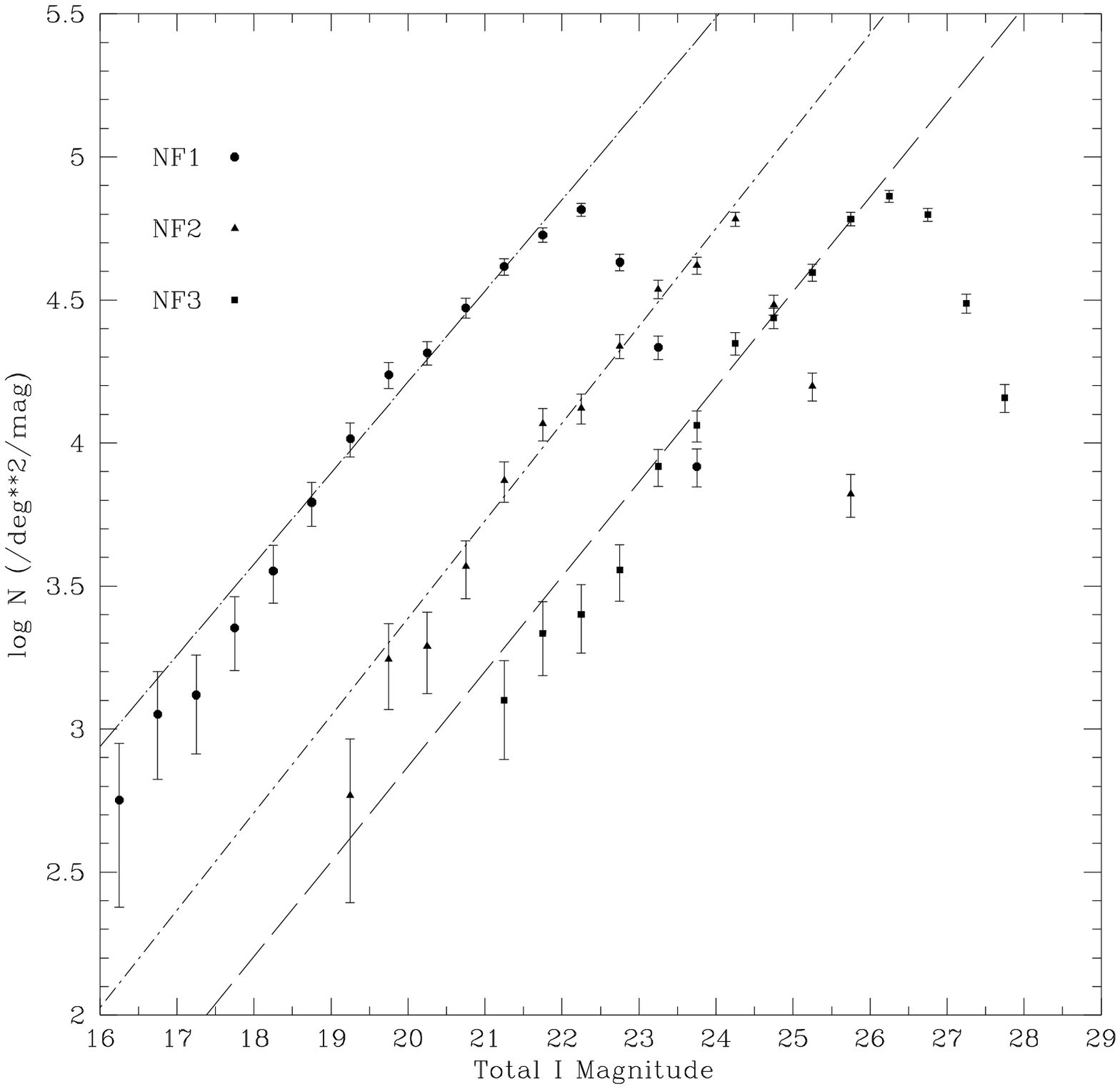]{$I$ number counts for NF1, NF2 and NF3.  
The plotted lines are least squares fits to all the points above the 
$I\sim24$ magnitude limits.  
Slopes are listed in Table 4 and the format of the plot
is the same as in Fig. 4. \label{fig6}}

\figcaption[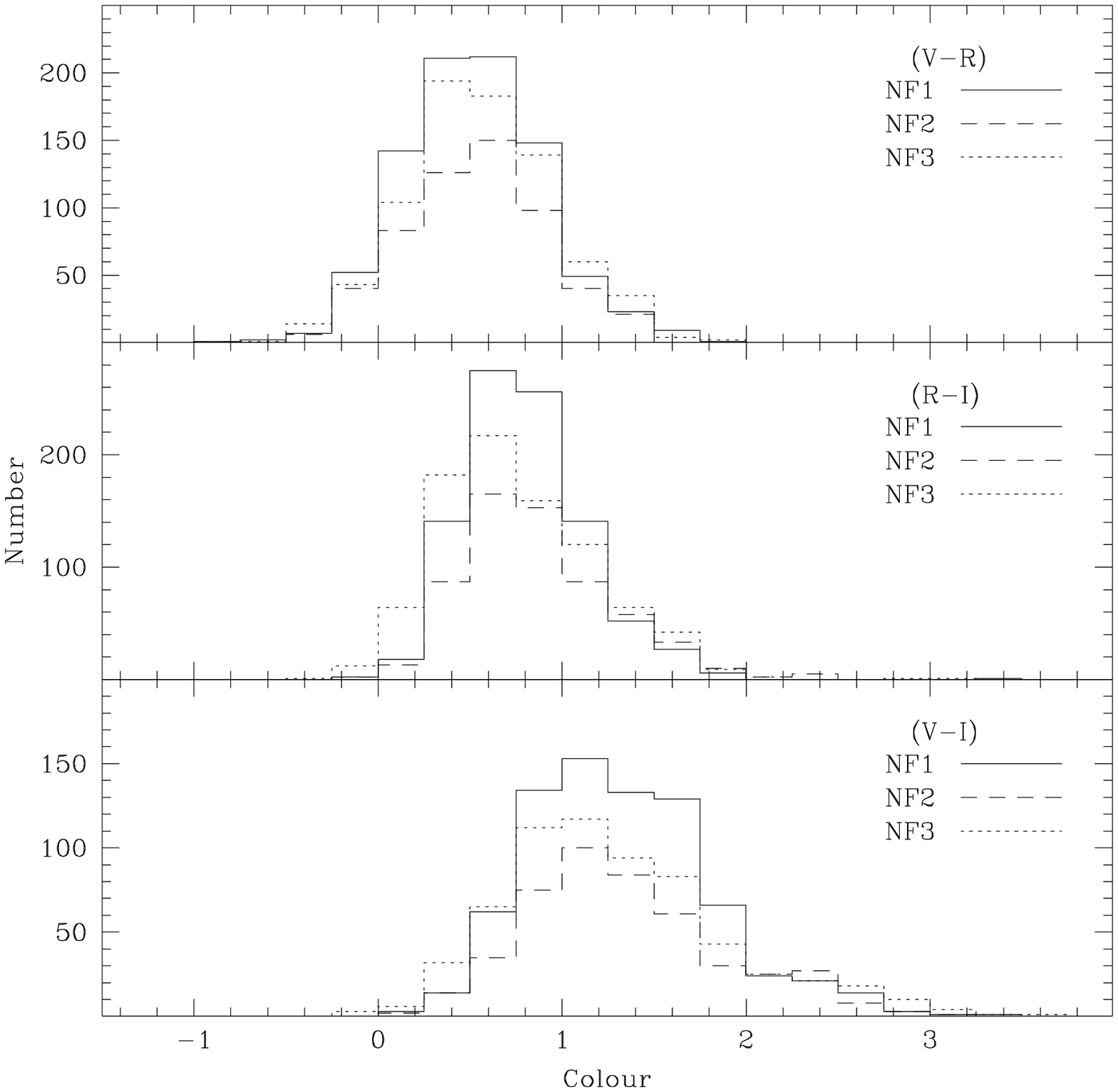]{Histograms of galaxy colors for NF1, NF2 and NF3 
with the number of galaxies observed as a function of $(V-R)$, 
$(R-I)$ and $(V-I)$ plotted in the top, middle and bottom panels, 
respectively.  The 
NF1, NF2 and NF3 fields are indicated by solid-line, 
dashed-line and dotted-line histograms. \label{fig7}}

\figcaption[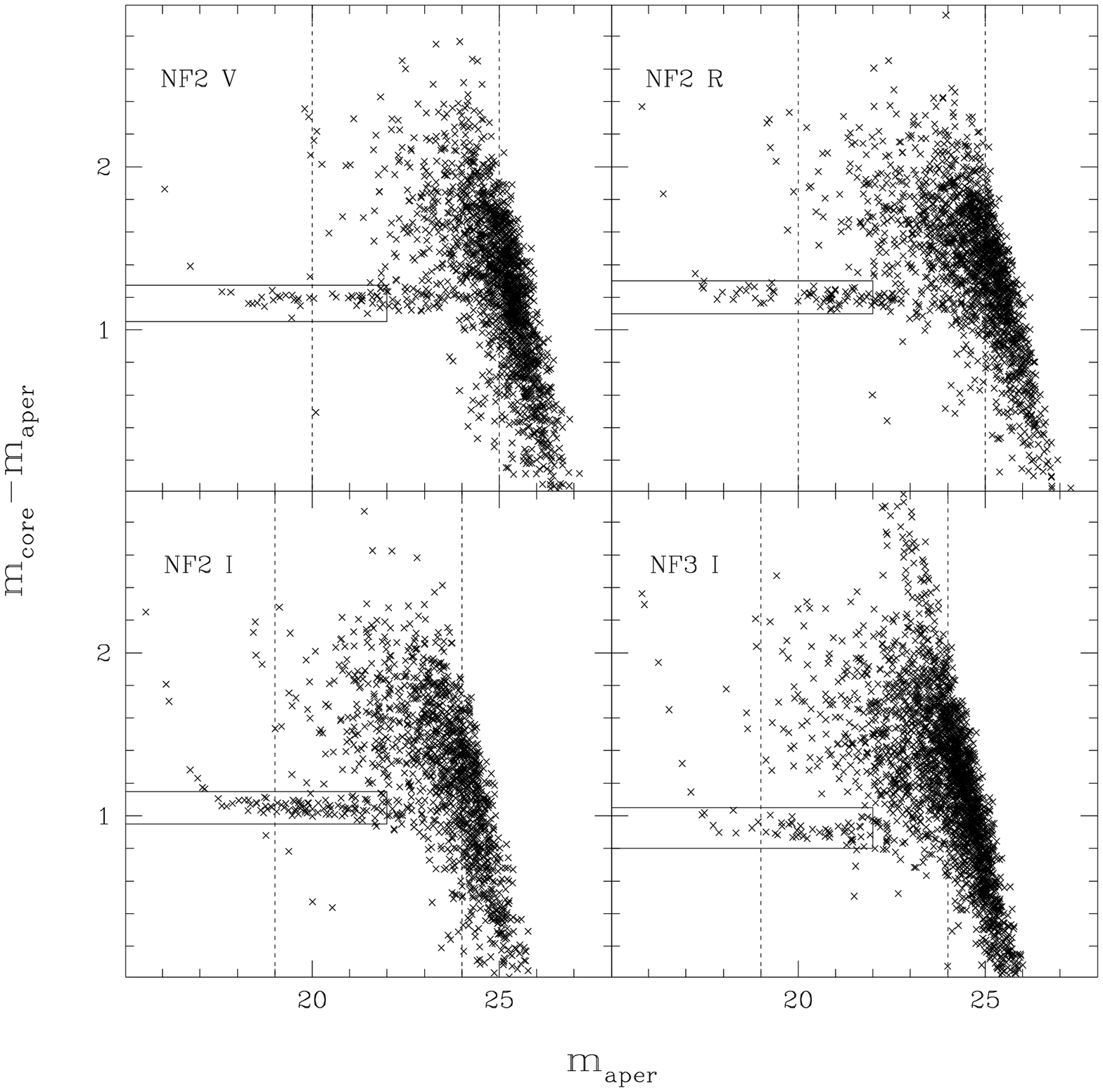]{Star-galaxy separation diagrams for NF2 and NF3 with
the difference between the ``core'' and aperture magnitudes plotted on the 
ordinate and just the latter on the abscissa.  The field and bandpass
are shown in the upper left corner of each panel.  Magnitude limits are 
marked with dashed lines and the stellar sequences are demarcated by 
solid line rectangular boxes.  Note the ``plume'' of galaxies seen 
in the NF3-$I$ panel, at the upper right of the data near the magnitude
limit, is caused by spurious noise objects yet to be removed from the 
galaxy catalog. \label{fig8}}

\figcaption[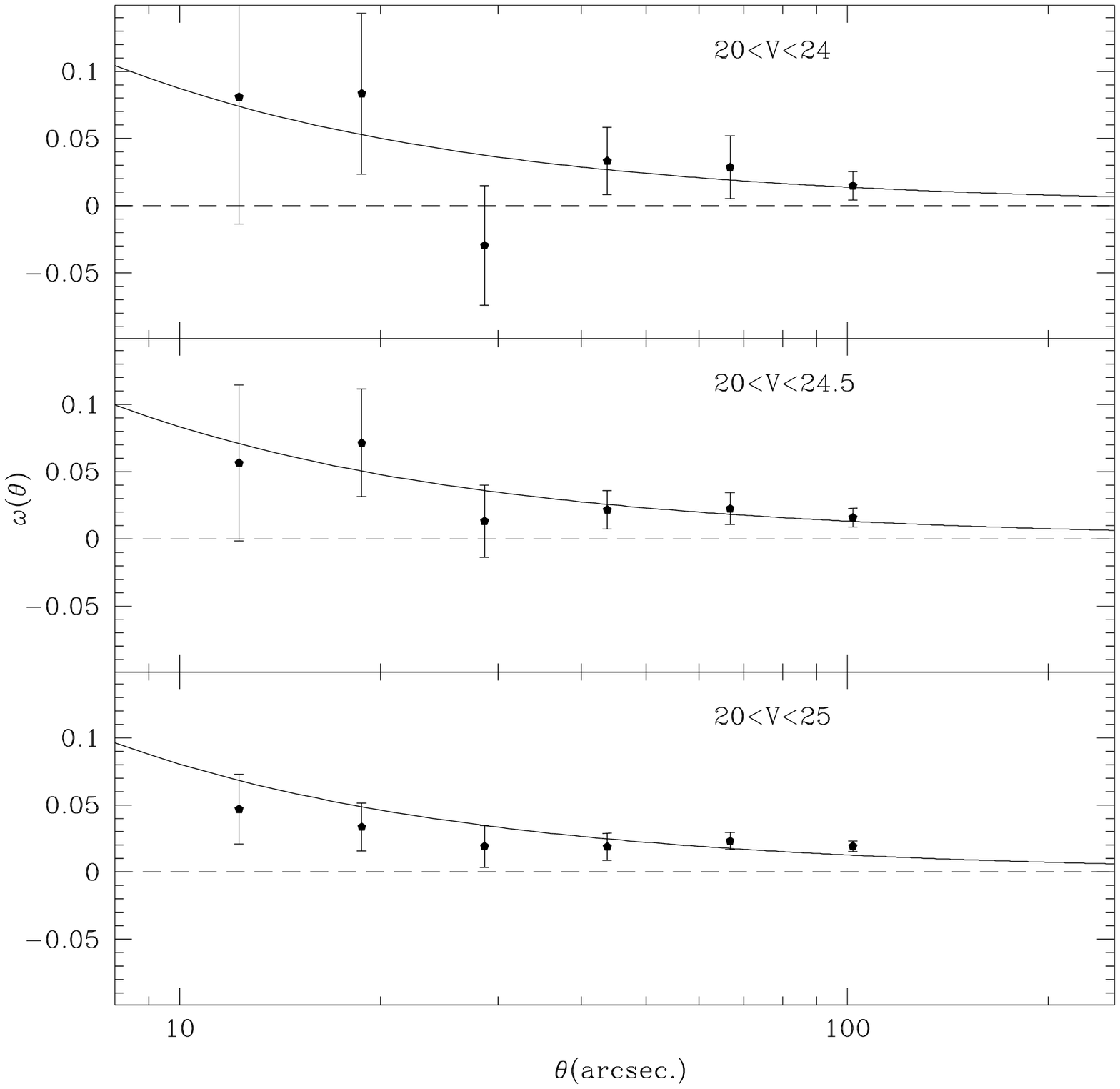]{Measurements of the angular correlation function 
using $V$-band data from NF1, NF2 and NF3, 
for the listed magnitude ranges.  
Error bars are calculated using bootstrap resampling. 
The solid lines represent the standard model fit to the data assuming 
$\delta=0.8$.  See text for details. \label{fig9}}

\figcaption[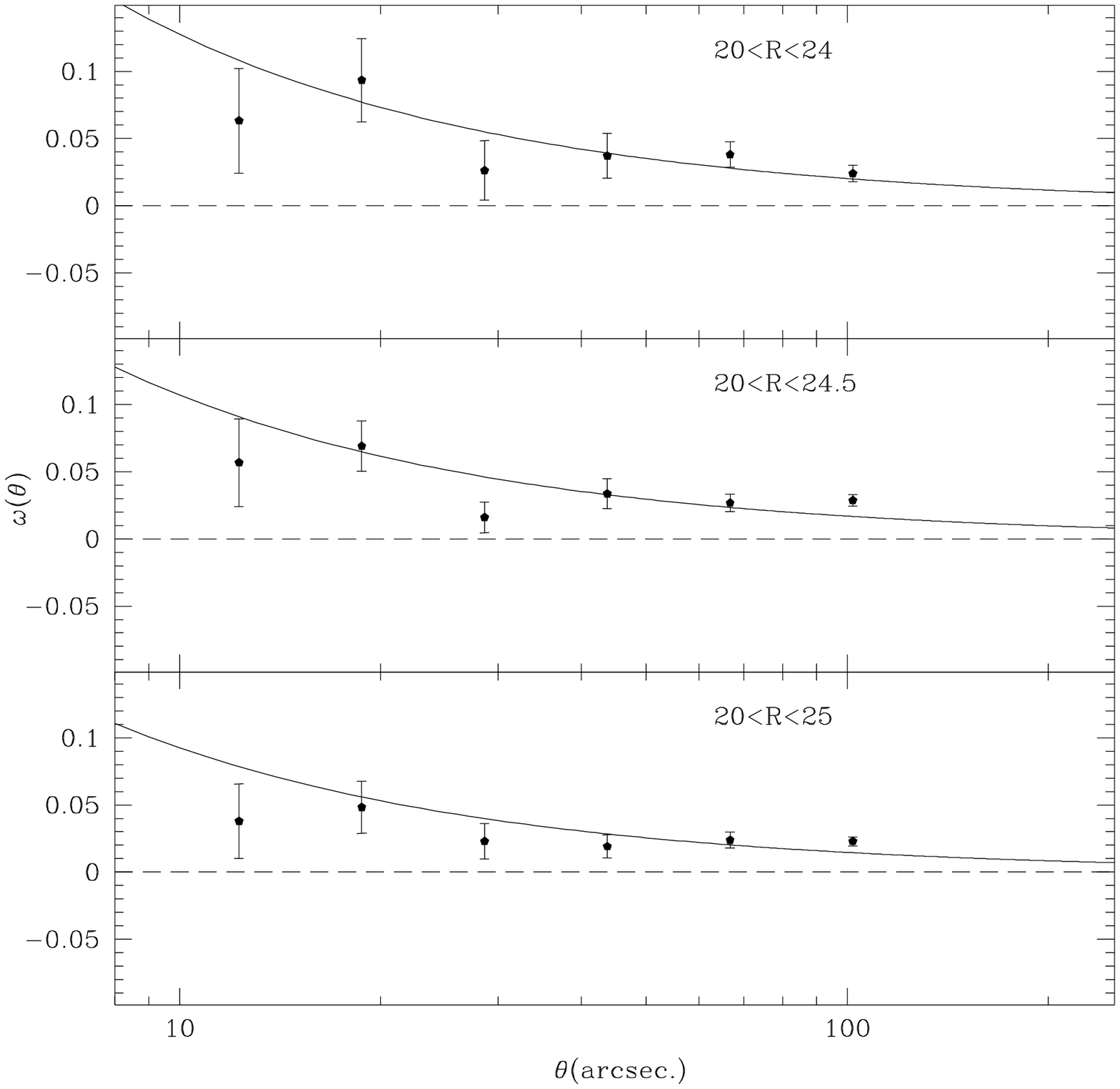]{As in Fig.9, but for the $R$ data. \label{fig10}}

\figcaption[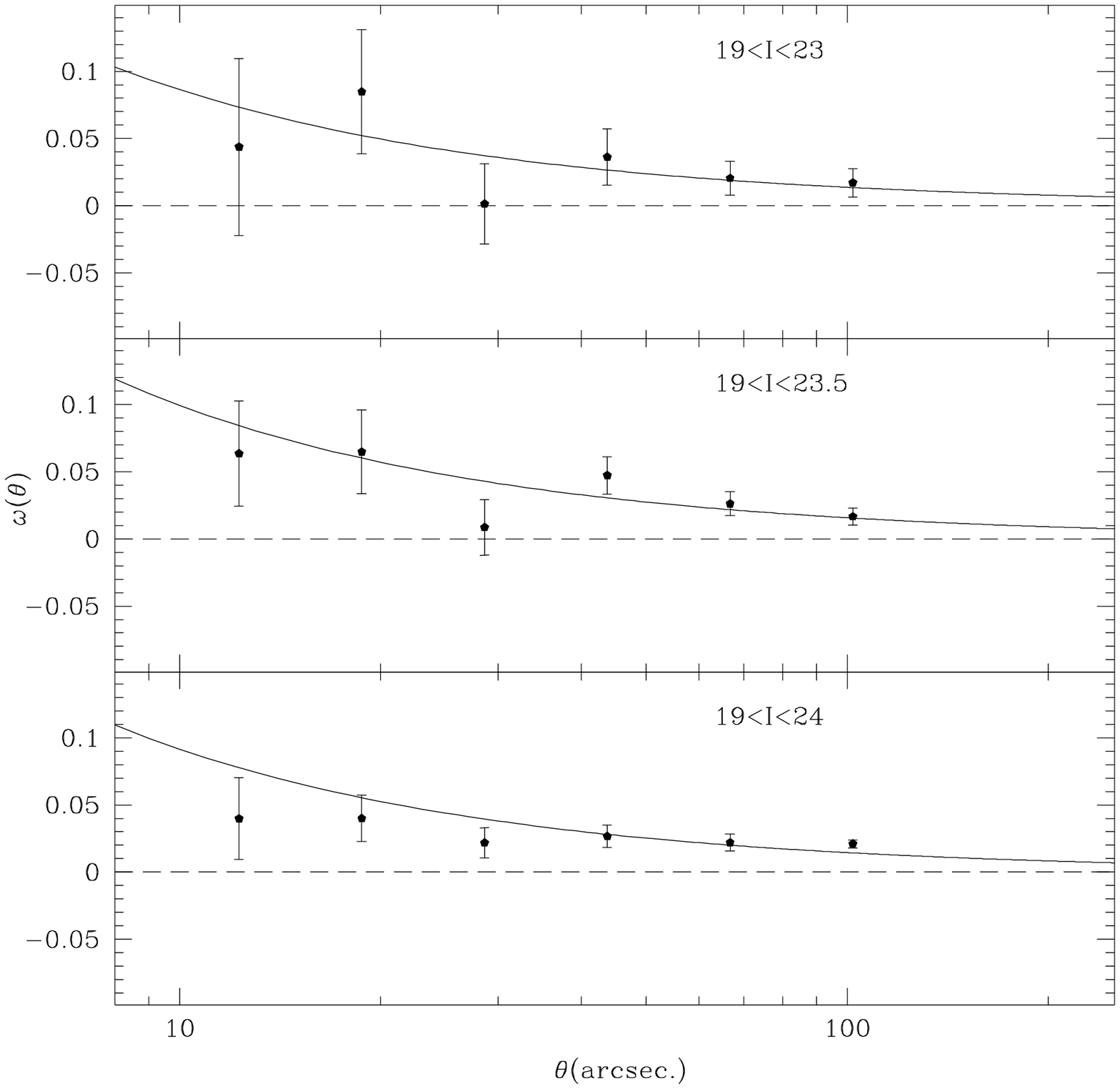]{As in Fig.9, but for the $I$ data. \label{fig11}}

\figcaption[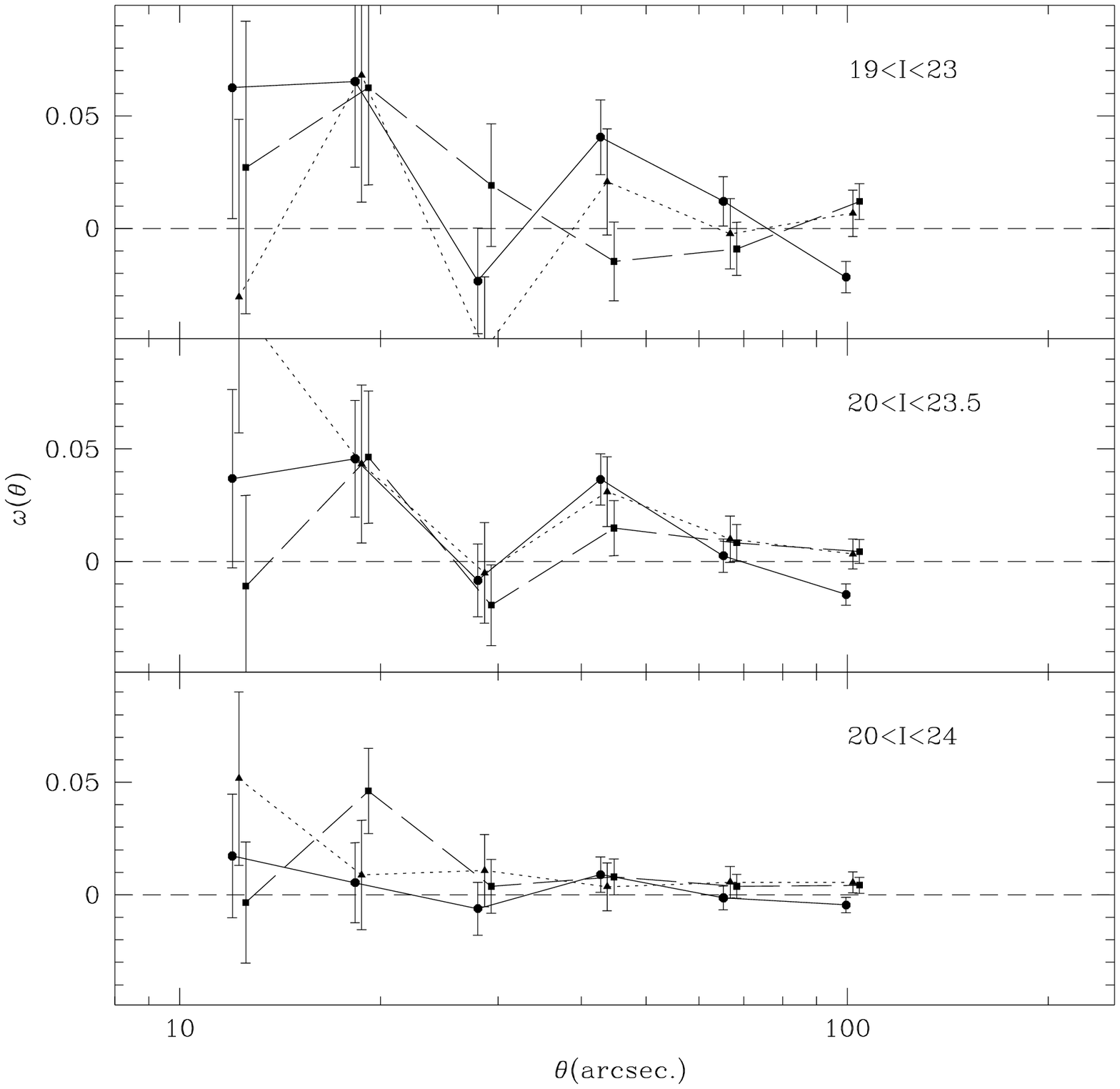]{Comparison of ``raw'' angular correlation functions 
measured for the listed magnitude ranges in each field, in the $I$-bandpass.  
The circles, triangles and squares denote the NF1, NF2 and NF3 data points, 
respectively.  NF1 data are offset $0.01$ dex to the left while the
NF3 data points have been shifted the same amount to the right, 
for the purpose of clarity.  These 
measurements have {\it not} been corrected for the integral constraint or 
for stellar dilution (see Table 7).  Better agreement is observed 
between the three fields at fainter magnitude limits, where the samples of galaxies 
are the largest.  Note the ordinate scale covers a smaller range than 
in Fig. 11. \label{fig12}}

\figcaption[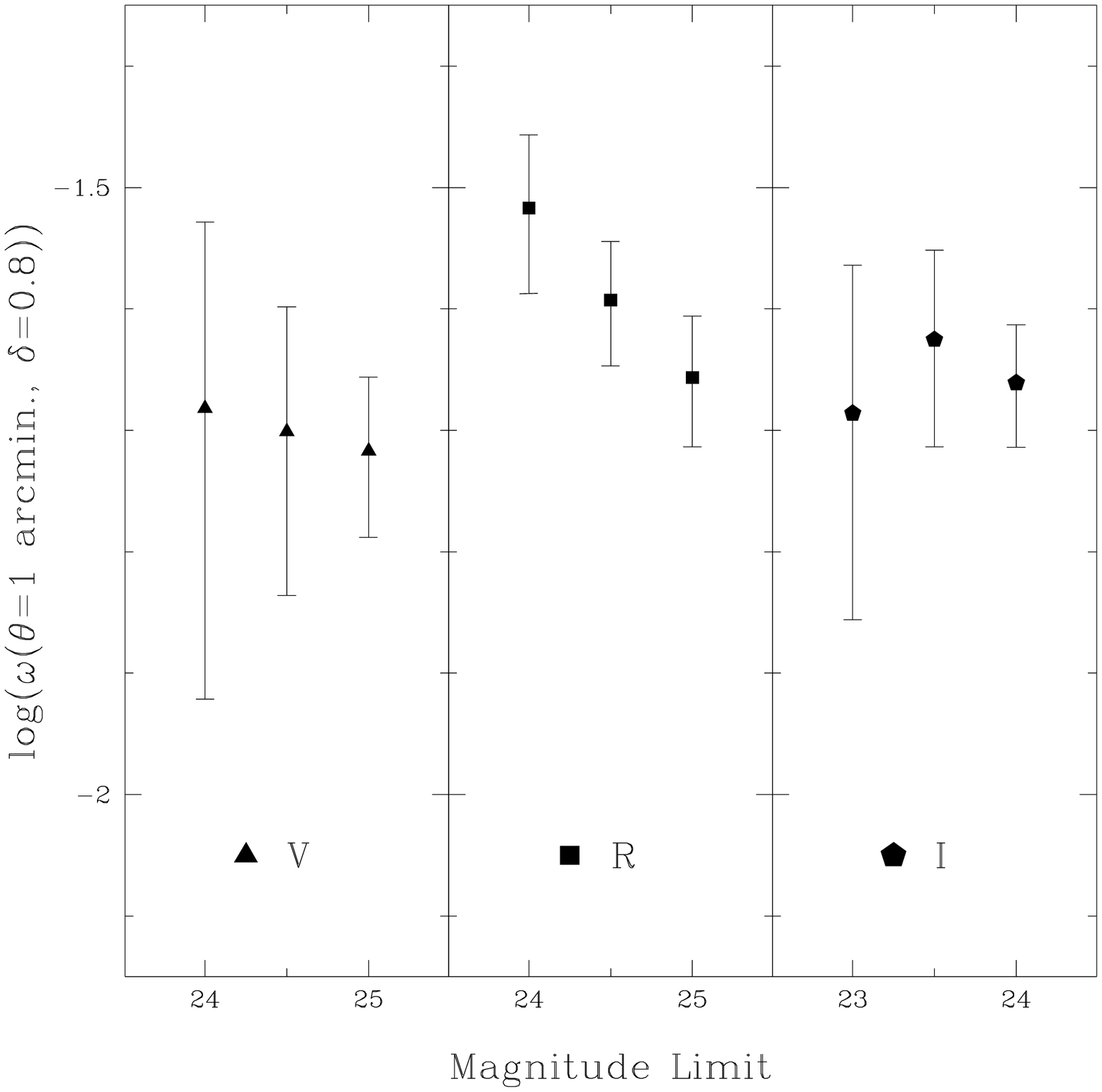]{The amplitudes of the angular correlation function 
calculated for a separation of one
arcminute for $V$, $R$ and $I$, assuming a power law with $\delta=0.8$.  
Measurements for the $V$, $R$ and  $I$-band data are shown in the 
left, center and right panels, respectively. \label{fig13}}

\figcaption[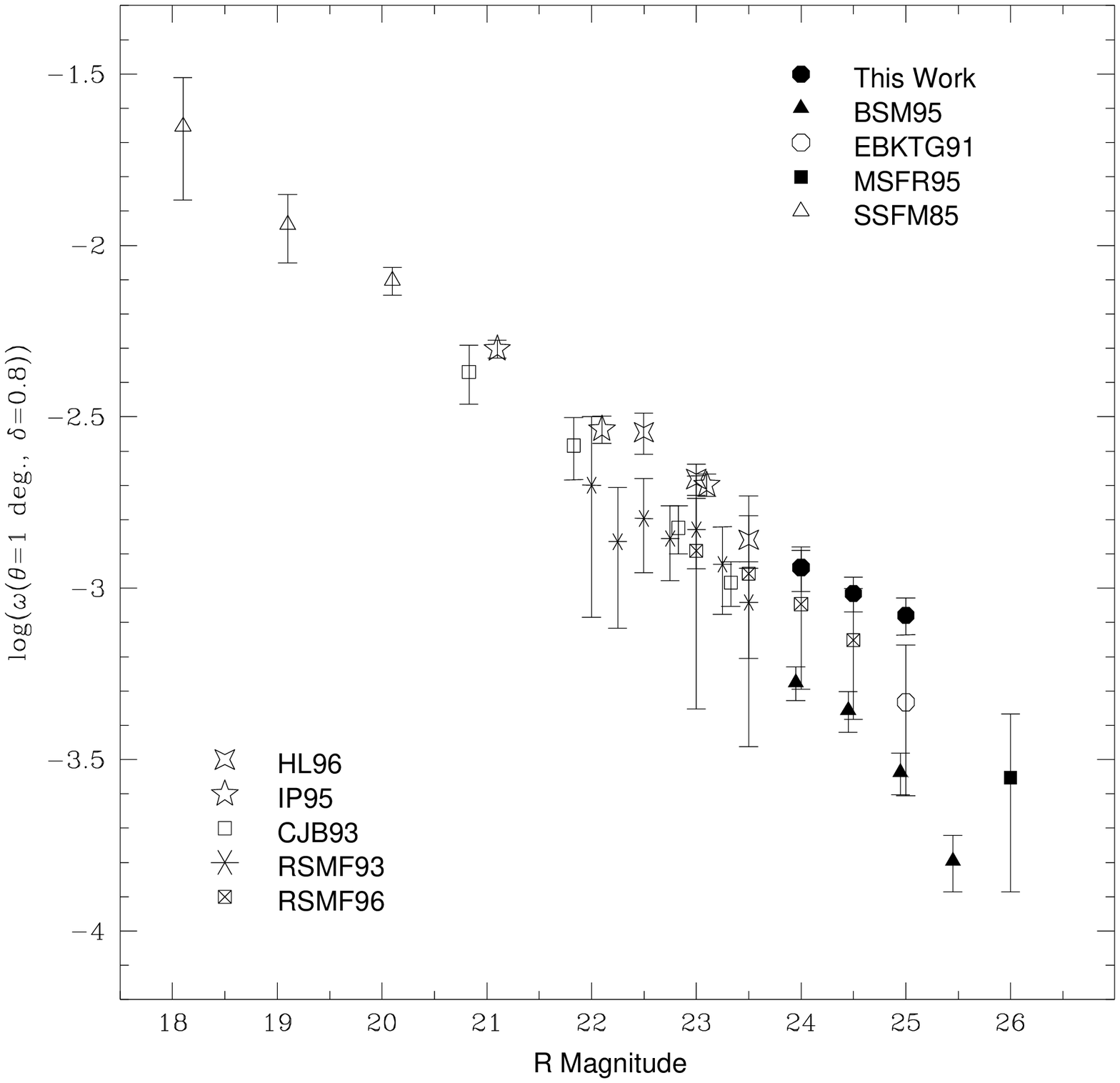]{Amplitudes of the angular correlation function 
normalized to 1 degree
assuming $\delta=0.8$ for this work and other
studies from the literature, in the $R$ band.  Details of the 
magnitude transformations used for the different observations 
are given in the text.  Each symbol is listed with the initials
of the authors' names and the year of the study it denotes. \label{fig14}}

\figcaption[colcf.eps]{Measurements of $\omega(\theta)$ for the 
full range of angular separations ($10-126''$; giving an effective separation 
of $\sim35''$) 
as a function 
of $(V-R)$, $(R-I)$ and $(V-I)$ colors.  The top panels show the 
clustering amplitudes as a function of colors the galaxies are {\it less}
than.  Similarly, the bottom panels show clustering with the abscissa
giving the colors the galaxies are {\it greater} than.  
Note the limited range of color the $V$, $R$ and $I$ filters provide.  
The Poisson error bars plotted are determined from the combination of the
three fields, and are probably underestimates of the true errors 
(see text). \label{fig15}}

\figcaption[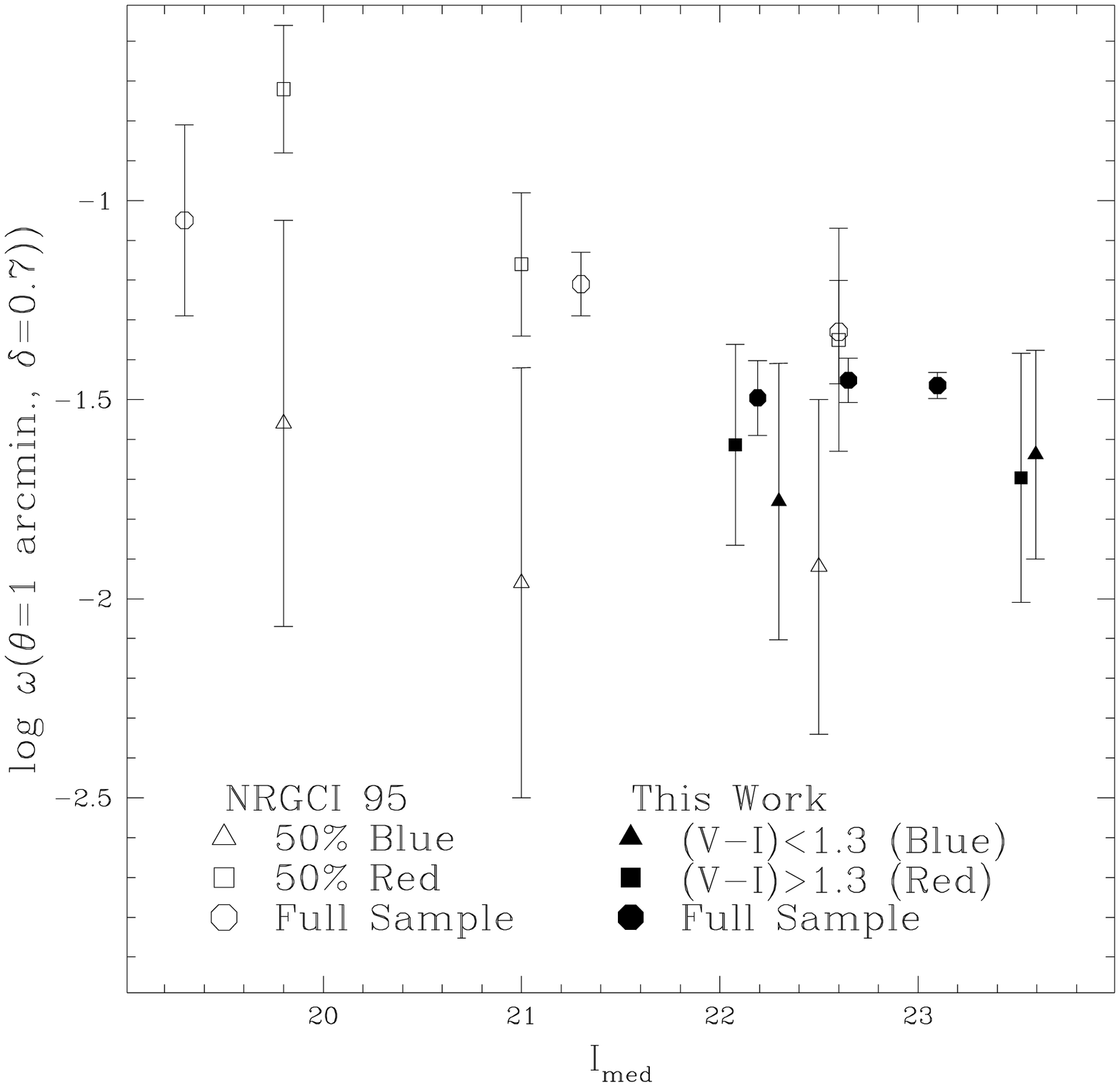]{Amplitudes of the angular correlation function 
calculated for separations of 1 arcminute 
assuming $\delta=0.7$, following Fig.3 of Neuschaefer {\it et al.} 
(1995, NRGCI)
to ease the comparison with this work.  The abscissa is the {\it median}
$I$ magnitude for each galaxy sample plotted.  Open symbols show the 
``50\% Blue'', ``50\% Red'' and full samples of NRGCI.  Filled symbols
show the red, blue and entire galaxy samples for this study, with 
$(V-I)=1.3$ used as the blue/red dividing line. \label{fig16}}

\figcaption[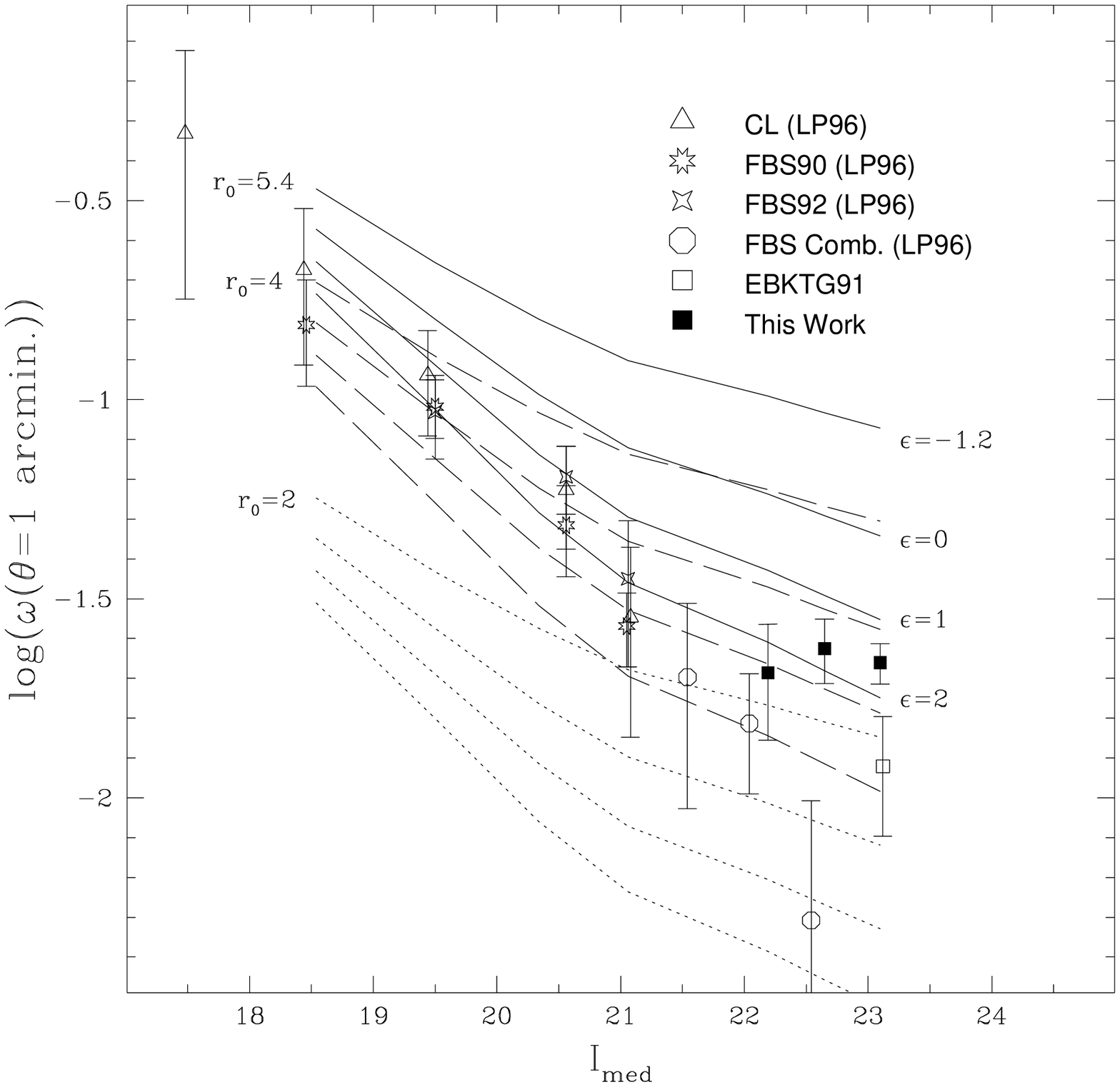]{The angular correlation function determined 
for separations of 1 arcminute plotted as a function of the {\it median} 
$I$ magnitude.  Results for 
the $I$-selected samples of LP, 
Efstathiou {\it et al.} (1991, EBKTG) and this work are shown.  Four different
subsamples of LP are listed with the corresponding symbols, along 
with the EBKTG observation and the filled squares which denote the 
$I$-filter measurements of this work.  Each family of lines correspond
to evolutionary models calculated for a given correlation length 
(solid lines: $r_{0}=5.4~h^{-1}~{\rm Mpc}$, 
dashed lines: $r_{0}=4.0~h^{-1}~{\rm Mpc}$ 
and dotted lines: $r_{0}=2.0~h^{-1}~{\rm Mpc}$) and 
$\epsilon=-1.2,0,1,2$ (from top to bottom, respectively). \label{fig17}}

\figcaption[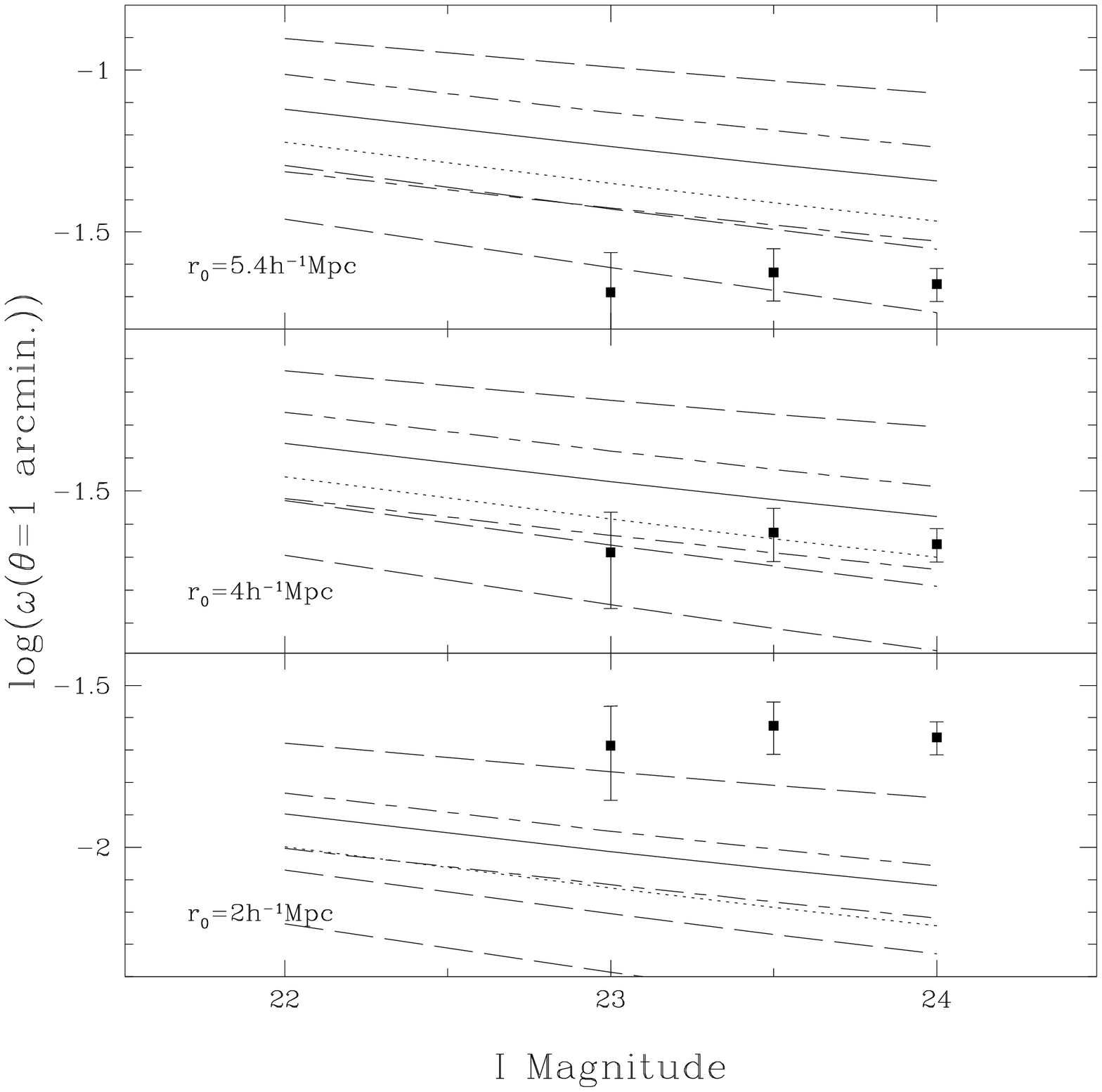]{A further comparison of the $I$-band observations 
of $\omega(\theta)$ with $\theta=1'$ to a suite of evolutionary models 
calculated using the extrapolated CFRS redshift distributions 
(Lilly {\it et al.} 1995a).  Note the abscissa is the $I$ magnitude 
limit {\it not} the median magnitude.  The correlation length ($r_{0}$)
used is shown in the lower left corner of each panel.  For each panel, 
the solid line corresponds to the model with $q_{0}=0.5$, $\gamma=1.8$
and $\epsilon=0$.  Just a change of $q_{0}=0.1$ yields the dotted line.   
Setting $\gamma=1.9,1.6$ moves the solid line to the short-long dashed 
lines above and below, respectively.  Changing just $\epsilon$, to -1.2, 
gives the long dashed line above the solid line, while the dashed lines
in decreasing amplitude below are for $\epsilon=1,2$. \label{fig18}}

\clearpage

\begin{table*}
\begin{center}
\begin{tabular}{cccccc}
\tableline
\tableline
Field & $\alpha_{1950}$ & $\delta_{1950}$ & \it{l}  & \it{b}  & Dates Observed \\
\tableline
NF1 & $13^{h}10^{m}10\fs80$ & $+43\arcdeg01\arcmin06\farcs0$ & $109.0\arcdeg$ & $+73.8\arcdeg$ & 1991 April 7-11 \\
NF2 & $15^{h}39^{m}01\fs50$ & $+24\arcdeg47\arcmin36\farcs0$ & $39.0\arcdeg$ & $+51.9\arcdeg$ & 1992 June 4-6, 1993 March 25\\
NF3 & $12^{h}29^{m}11\fs70$ & $+02\arcdeg07\arcmin42\farcs0$ & $291.5\arcdeg$ & $+64.3\arcdeg$ & 1993 March 23-25 \\
\tableline
\end{tabular}
\end{center}
\caption{Coordinates and Observing Runs of ``Blank Fields''} \label{tbl-1}
\end{table*}

\clearpage

\begin{table*}
\begin{center}
\begin{tabular}{ccccccccc}
\tableline
\tableline
\multicolumn{3}{c}{Filters} &\multicolumn{2}{c}{NF1} &\multicolumn{2}{c}{NF2} & \multicolumn{2}{c}{NF3}  \\
 &   \multicolumn{1}{c}{$\lambda_{eff}$(\AA)} &\multicolumn{1}{c}{$\Delta\lambda$(\AA)} & Exp. (s) & Seeing  & Exp. (s) & Seeing & Exp. (s) & Seeing  \\
\tableline
$V$ &  5430&  900& 9600 & $0\farcs94$ &  9000 & $0\farcs82$ & 4800 & $0\farcs72$ \\
$R$ &  6485&  1280& 9600 & $0\farcs89$ &  6300 & $0\farcs84$ & 6000 & $0\farcs66$ \\
$I$ &  8320&  1950& 12000 & $0\farcs83$ &  6000 & $0\farcs76$ & 8400 & $0\farcs73$ \\
\tableline
\end{tabular}
\end{center}
\caption{Filters, Total Exposure Times and Average Seeing} \label{tbl-2}
\end{table*}

\clearpage

\begin{table*}
\begin{center}
\begin{tabular}{cc}
\tableline
\tableline
    \multicolumn{1}{c}{Observing Run} &\multicolumn{1}{c}{Standards} \\
\tableline
 1991 April & M67 (Montgomery {\it et al.} \markcite{M93}1993, Schild \markcite{S83}1983), \\
  & NGC 4147 (Christian {\it et al.} \markcite{C85}1985 [C85])\\
 1992 June &  NGC 4147 (C85), SA 110 (Landolt \markcite{L92}1992 [L92]), \\
  & SA 113 (L92), M92 (L. Davis, priv. comm. [D])\\
 1993 March & SA 98 (L92), M92 (D),\\
  & G12-43 (L92), RU 149 (L92)\\
\tableline
\end{tabular}
\end{center}
\caption{Standard Star Fields} \label{tbl-3}
\end{table*}

\clearpage

\begin{table*}
\begin{center}
\begin{tabular}{cccc}
\tableline
\tableline
Field & $V$ & $R$ & $I$ \\
\tableline
NF1 & $0.41\pm0.01$ & $0.36\pm0.01$ & $0.32\pm0.01$  \\
NF2 & $0.42\pm0.02$ & $0.35\pm0.01$ & $0.34\pm0.02$  \\
NF3 & $0.46\pm0.02$ & $0.39\pm0.02$ & $0.33\pm0.02$  \\
\tableline
\end{tabular}
\end{center}
\caption{Slopes of Number Counts} \label{tbl-4}
\end{table*}

\clearpage

\begin{table*}
\begin{center}
\begin{tabular}{cccc}
\tableline
\tableline
\multicolumn{4}{c}{NF1 ({\it Effective Area}: $\sim0.01064~deg^{2}$)}  \\
Filter & \multicolumn{3}{c}{Number of Objects (Magnitude Range)}\\
\tableline
$V$ & 355 (20-24) & 590 (20-24.5) & 935 (20-25) \\
$R$ & 574 (20-24) & 878 (20-24.5) & {...} \\
$I$ & 486 (19-23) & 706 (19-23.5) & 996 (19-24) \\
\tableline
\multicolumn{4}{c}{NF2 ({\it Effective Area}: $\sim0.01026~deg^{2}$)}  \\
Filter & \multicolumn{3}{c}{Number of Objects (Magnitude Range)}\\
\tableline
$V$ & 253 (20-24) & 425 (20-24.5) & 699 (20-25) \\
$R$ & 407 (20-24) & 610 (20-24.5) & 913 (20-25) \\
$I$ & 319 (19-23) & 496 (19-23.5) & 709 (19-24) \\
\tableline
\multicolumn{4}{c}{NF3 ({\it Effective Area}: $\sim0.01111~deg^{2}$)}  \\
Filter & \multicolumn{3}{c}{Number of Objects (Magnitude Range)}\\
\tableline
$V$ & 312 (20-24) & 543 (20-24.5) & 908 (20-25) \\
$R$ & 588 (20-24) & 930 (20-24.5) & 1482 (20-25) \\
$I$ & 439 (19-23) & 658 (19-23.5) & 992 (19-24) \\
\tableline
\end{tabular}
\end{center}
\caption{Number of Objects Detected } \label{tbl-5}
\end{table*}

\clearpage

\begin{table*}
\begin{center}
\begin{tabular}{ccccccc}
\tableline
\tableline
  & \multicolumn{2}{c}{$20\leq V \leq25$} & \multicolumn{2}{c}{$20\leq R \leq25$}& \multicolumn{2}{c}{$19\leq I \leq24$} \\
\tableline
  & $A_{\omega}^{fin}$ & $\chi^{2}$ & $A_{\omega}^{fin}$ & $\chi^{2}$ & $A_{\omega}^{fin}$ & $\chi^{2}$ \\
\tableline
$\delta=0.5$ & $0.653\pm0.024$ & 49.3 & $0.686\pm0.022$ & 62.6 & $0.689\pm0.020$ & 67.5 \\
$\delta=0.6$ & $0.596\pm0.036$ & 26.9 & $0.640\pm0.033$ & 36.7 & $0.641\pm0.031$ & 37.9 \\
$\delta=0.7$ & $0.546\pm0.053$ & 13.7 & $0.605\pm0.049$ & 20.9 & $0.603\pm0.045$ & 20.2 \\
$\delta=0.8$ & $0.508\pm0.077$ & 6.6 & $0.583\pm0.072$ & 11.9 & $0.578\pm0.067$ & 10.5 \\
$\delta=0.9$ & $0.485\pm0.110$ & 3.1 & $0.581\pm0.104$ & 7.1 & $0.570\pm0.097$ & 5.4 \\
\tableline
\end{tabular}
\end{center}
\caption{Angular Correlation Function Fits and $\chi^{2}$ Statistic (five degrees of freedom)} \label{tbl-6}
\end{table*}

\clearpage

\begin{table*}
\begin{center}
\begin{tabular}{cccc}
\tableline
\tableline
 $\bf{V}$ &  $20\leq V \leq24$  & $20\leq V \leq24.5$ & $20\leq V \leq25$ \\
\tableline
$A_{\omega}^{fin}$ & $0.550\pm0.233$ & $0.527\pm0.141$ & $0.508\pm0.077$ \\
$\left(\frac{N_{obj}}{N_{obj}-N_{s}}\right)^{2}$ (NF1) & 1.116 & 1.087 & 1.067 \\
$~~~~~~''~~~~~~~$ (NF2) & 1.494 & 1.356 & 1.255 \\
$~~~~~~''~~~~~~~$ (NF3) & 1.198 & 1.147 & 1.104 \\
\tableline
 $\bf{R}$ &  $20\leq R \leq24$  & $20\leq R \leq24.5$ & $20\leq R \leq25$ \\
\tableline
$A_{\omega}^{fin}$ & $0.804\pm0.120$ & $0.676\pm0.079$ & $0.583\pm0.072$ \\
$\left(\frac{N_{obj}}{N_{obj}-N_{s}}\right)^{2}$ (NF1) & 1.105 & 1.082 & {...} \\
$~~~~~~''~~~~~~~$ (NF2) & 1.300 & 1.253 & 1.204 \\
$~~~~~~''~~~~~~~$ (NF3) & 1.156 & 1.117 & 1.085 \\
\tableline
 $\bf{I}$ &  $19\leq I \leq23$  & $19\leq I \leq23.5$ & $19\leq I \leq24$ \\
\tableline
$A_{\omega}^{fin}$ & $0.545\pm0.176$ & $0.627\pm0.116$ & $0.578\pm0.067$ \\
$\left(\frac{N_{obj}}{N_{obj}-N_{s}}\right)^{2}$ (NF1) & 1.083 & 1.075 & 1.065 \\
$~~~~~~''~~~~~~~$ (NF2) & 1.185 & 1.183 & 1.179 \\
$~~~~~~''~~~~~~~$ (NF3) & 1.072 & 1.077 & 1.070 \\
\tableline
\end{tabular}
\end{center}
\caption{Angular Correlation Function Amplitudes and Stellar Dilution Corrections} \label{tbl-7}
\end{table*}

\clearpage

\begin{table*}
\begin{center}
\begin{tabular}{lccc}
\tableline
\tableline
Model & $r_{0}~(h^{-1}~{\rm Mpc})$ & $\epsilon$ & Comments \\
\tableline
(1) Bursting Dwarfs & $\sim4$ & $\sim0-1$ & acceptable at all limits; only moderate \\
 & & & clustering evolution required  \\
(2) Fading Dwarfs & $\sim2$ & $\sim-1.2$ & untestable locally (?) \\
(3) Merging & $\sim5$ & $\sim1-2$ & excessive (unobserved) clustering \\
 & & & evolution may be required \\
\tableline
\end{tabular}
\end{center}
\caption{Summary of Galaxy Evolution Models} \label{tbl-8}
\end{table*}

 
 
 
\clearpage
\pagestyle{empty}
\plotone{vcount.eps}
 
\clearpage
\plotone{rcount.eps}
 
\clearpage
\plotone{icount.eps}
 
\clearpage
\plotone{colhist.eps}
 
\clearpage
\plotone{sgsep2.eps}
 
\clearpage
\plotone{vwth.eps}
 
\clearpage
\plotone{rwth.eps}
 
\clearpage
\plotone{iwth.eps}

\clearpage
\plotone{icfin.eps}
 
\clearpage
\plotone{colcomp.eps}
 
\clearpage
\plotone{compw.eps}
 
\clearpage
\plotone{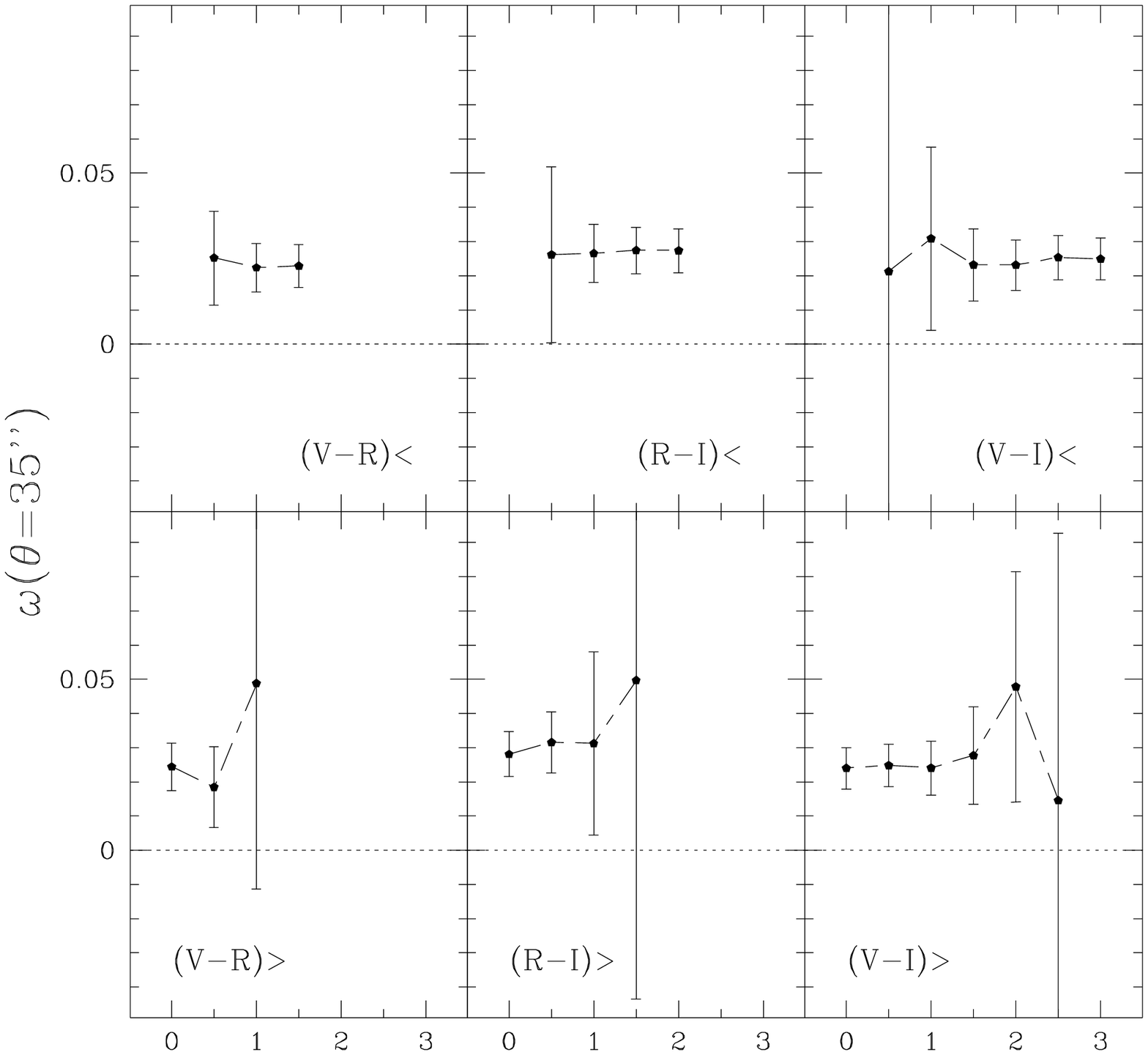}
 
\clearpage
\plotone{neucomp.eps}
 
\clearpage
\plotone{compimod.eps}
 
\clearpage
\plotone{imod.eps}

\end{document}